\documentclass[rmp,aps,nofootinbib,twocolumn]{revtex4}

\usepackage{graphics}
\usepackage{epsfig}
\usepackage{bm}
\usepackage{epsfig}
\usepackage{graphicx}
\usepackage{amsmath}
\usepackage{amssymb}
\usepackage{color}
\usepackage{euscript}
\usepackage{mathrsfs}

\newcommand{\ave}[1]{{\langle #1\rangle}}

\newcommand{\bra}[1]{{\langle #1 \vert}}

\newcommand{\ket}[1]{{\vert #1 \rangle}}

\begin{document}
\title{Colloquium:
Fundamental aspects of steady state heat to work conversion}

\author{Giuliano Benenti}
\email{giuliano.benenti@uninsubria.it}
\affiliation{CNISM \& Center for Nonlinear and Complex Systems,
Universit\`a degli Studi dell'Insubria, Via Valleggio 11, 22100 Como, Italy}
\affiliation{Istituto Nazionale di Fisica Nucleare, Sezione di Milano,
via Celoria 16, 20133 Milano, Italy}
\author{Giulio Casati}
\email{giulio.casati@uninsubria.it}
\affiliation{CNISM \& Center for Nonlinear and Complex Systems,
Universit\`a degli Studi dell'Insubria, Via Valleggio 11, 22100 Como, Italy}
\affiliation{Istituto Nazionale di Fisica Nucleare, Sezione di Milano,
via Celoria 16, 20133 Milano, Italy}
\author{Toma\v z Prosen}
\email{tomaz.prosen@fmf.uni-lj.si}
\affiliation{Department of Physics, Faculty of Mathematics and Physics,
University of Ljubljana, Ljubljana, Slovenia.}
\author{Keiji Saito}
\email{saitoh@rk.phys.keio.ac.jp}
\affiliation{Department of Physics, Keio University
3-14-1 Hiyoshi, Kohoku-ku, Yokohama 223-8522, Japan}

\begin{abstract}
We review theoretical approaches to analyzing efficiency of steady state heat to work conversion which is crucial in the timely problem of optimizing efficiency of small-scale heat engines and refrigerators. A rather abstract perspective of  non-equilibrium statistical mechanics and dynamical system's theory is taken to view at this very practical problem. Several recently discovered general mechanisms of optimizing the figure of merit of thermoelectric efficiency are discussed, also 
in connection to breaking time-reversal symmetry of the microscopic equations of motion. Applications of these theoretical and mathematical ideas to practically relevant models are pointed out.
\end{abstract}

\date{\today}
\maketitle
\tableofcontents

\section{INTRODUCTION}
\label{sec:intro}

The need of providing a sustainable energy to the world population is becoming increasingly important. It is likely that in the following decades the efforts of the scientific community will be increasingly addressed to this direction and in particular to the heat to work transformation. An important possibility under investigation is the thermoelectric power generation and refrigeration. In spite of relevant progress made in the last years the efficiency of thermoelectric technology remains too low \cite{goldsmid,dresselhaus,snyder,kanatzidis,majumdarrev,shakouri,dubi}.
Indeed such efficiency depends on physical properties of a given material namely the electrical conductivity $\sigma$, the thermal conductivity $\kappa$, and the Seebeck coefficient $S$, and is expressed by a non-dimensional quantity, often called figure of merit, 
$ZT=(\sigma S^2/\kappa)\, T$ \cite{ioffebook}. 
High efficiency requires high $ZT$ values which seem difficult to achieve.  After more than 50 years from Ioffe's discovery that doped semiconductors exhibit a relatively large thermoelectric effect
\cite{ioffebook,ioffereview}, 
and in spite of recent achievements, the most efficient actual devices still operate at $ZT$ around 1. Certainly, in consideration of the importance of the problem,  even a small improvement would be most welcome. However, it is generally accepted that $ZT\approx 3$ is a target value for efficient competing thermoelectric technology and, so far, no clear paths exist which may lead to reach that target. 

In such a situation it is probably useful to investigate a different approach namely an approach which starts from first principles i.e. from the fundamental microscopic dynamical mechanisms which determine the phenomenological laws of heat and particles transport. In this connection, as it is well known, the enormous achievements in nonlinear dynamical systems and the new tools developed have led to a much better understanding of the statistical behavior of dynamical systems. For example, the question of the derivation of the phenomenological Fourier law of heat conduction from the dynamical equations of motion has been studied in great detail \cite{lepri,dhar}. Theoretical work in this direction even led to the possibility to control the heat current and devise heat diodes, 
transistors, and thermal logic gates
\cite{hanggi2012}. Preliminary experimental results have also been 
obtained \cite{majumdar,terasaki}.  We are confident that this theoretical approach, combined with the present sophisticated numerical techniques, may lead to substantial progress on the way of improving the long standing problem of thermoelectric efficiency. An additional motivation in favor of this approach is that thermoelectric technology, at small sizes (e.g. at micro or nano-scale), is expected to be more efficient than traditional conversion systems. Indeed the efficiency of mechanical engines decrease very rapidly at low power level. The recent progress in engineering nanostructured materials opens now new possibilities. The study of dynamical complexity of these structures may lead to the design of new strategies for developing  materials with high thermoelectric efficiency. Nanostructures may allow to control the thermal and electrical conductivity e.g. with appropriate 
scattering mechanisms \cite{dresselhaus,majumdarrev,shakouri}.

In summary, what is required is a better understanding of the fundamental dynamical mechanisms which control heat and particles transport. The combined efforts of physicists and mathematicians working in nonlinear dynamical systems and statistical mechanics, condensed matter physicists, and material scientists may prove useful to contribute substantially to the progress in this field of great importance for both energy supply and the environmental concern. 
 
The purpose of the present Colloquium is to introduce the basic tools and fundamental results on steady state heat to work conversion, mainly from a rather abstract, statistical physics and dynamical system's perspective, yet with a clear focus toward potential applications. We hope our paper might help bridging the gap among rather diverse communities and research fields, such as non-equilibrium statistical mechanics, mathematical physics and dynamical systems, mesoscopic physics, and strongly correlated many-body systems of condensed matter. Our line of presentation is going from more abstract to more phenomenological.
We start with a short overview of non-equilibrium thermodynamics in section \ref{sec:nonequilibrium}, where fundamental results on linear response theory and Onsager reciprocity relations are discussed. In section \ref{sec:TE}
we then explain basic abstract definitions of thermoelectric heat to work conversion efficiency. In section \ref{sec:landauer} we review the microscopic Landauer-B\"uttiker framework for non-interacting systems and discuss the main non-interacting concepts for controlling thermodynamic efficiency: energy filtering, external noise and probe reservoirs. We believe that most of exciting future investigations on heat to work conversion will be devoted to strongly interacting systems, therefore we set the stage in section \ref{sec:interacting} by reviewing state of the art on understanding thermalization in equilibrium and local-thermalization near-equilibrium in closed and open interacting systems.  As the analysis of strongly interacting systems is mainly relying on numerical simulations, we outline in section \ref{sec:numerics} some of the key ideas and methods for efficient simulation of non-equilibrium steady states of classical and quantum open  many-body systems. In section \ref{sec:machines}  we then discuss some simple models of thermoelectric engines and stress their importance from either 
exact-solvability or practical-relevance perspective. In section \ref{sec:phenomenology} we outline some phenomenological and empirical laws governing thermoelectric phenomena, with the emphasis on open theoretical problems.
We conclude in Sec.~\ref{sec:conclusions} with some 
remarks on future prospects of the field.

\section{OVERVIEW OF BASIC CONCEPTS OF NON-EQUILIBRIUM THERMODYNAMICS}
\label{sec:nonequilibrium}

Non-equilibrium thermodynamics \cite{callen,mazur} describes processes
on the basis of two types of parameters:
\emph{thermodynamic forces} $X_i$ (also known as generalized forces
or affinities) driving irreversible processes, and the
\emph{fluxes} $J_i$
characterizing the response of the system to the applied forces.
More specifically, we will consider a generic setup for the
extraction of work from a heat flow. The system performs
work $W=-F x$ against an external force $F$, with thermodynamically
conjugate variable $x$. The force can be of mechanical,
chemical, or electrical nature.
The thermodynamic force is $X_1=F/T$, with $T$
being the temperature of the system.
The thermodynamic flux is $J_1=\dot{x}$, where the dot denotes the
time derivative.
The output power reads $P=\dot{W}=-J_1X_1T$.
We are considering heat to work conversion,
that is, the work is performed by converting a part of the
amount of heat $Q_1$ flowing from the hot reservoir at temperature
$T_1$ (we assume $T_1>T_2$).
The thermodynamic force is $X_2=1/T_2-1/T_1$, and
the \emph{heat current} reads $J_2=\dot{Q}_1$.

\begin{figure}
\begin{center}
\epsfxsize=80mm\epsffile{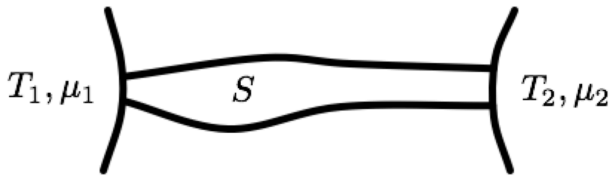}
\caption{Schematic drawing of steady-state thermoelectric heat to work
conversion. A system $S$ is in touch with two reservoirs at temperatures
$T_1,T_2$ and electrochemical potentials $\mu_1,\mu_2$. We assume
$T_1>T_2$ and $\mu_1<\mu_2$, so that the generalized forces $X_1<0$
and $X_2>0$. In thermoelectric power generation, both charge and 
heat flow from the hot to the cold reservoir, i.e., $J_1>0$ and
$J_2>0$. In refrigeration, $J_1<0$ and $J_2<0$. Note that, while
fluxes are one dimensional (along the direction connecting
the two reservoirs), the motion of particles or other degrees of freedom inside the system can be two or
three dimensional.} 
\label{fig:scheme}
\end{center}
\end{figure}

For instance, in thermoelectric power generation
(see Fig.~\ref{fig:scheme})
$F=\Delta V=\Delta\mu/ e$,
where $e$ is the electron charge and $\Delta V=V_1-V_2$
($V_1<V_2$)
the voltage difference between the two reservoirs at
electrochemical potentials $\mu_1$ and $\mu_2$,
$x$ is the total charge transferred from reservoir 1 to reservoir 2,
$X_1=\Delta V/T$, and $J_1$ is the steady-state \emph{electric current}.
We can also write $J_1=e J_\rho$, with $J_\rho$ being the particle current. 

\subsection{Linear response, Onsager reciprocal relations}
\label{sec:linear_response}

Assuming that the generalized forces are small, the relationship
between fluxes and forces is linear:
\begin{eqnarray}
\left\{
\begin{array}{l}
J_1=L_{11} X_1 + L_{12} X_2,
\\
\\
J_2=L_{21} X_1 + L_{22} X_2.
\end{array}
\right.
\label{eq:coupledlinear}
\end{eqnarray}
These relations are referred to as phenomenological
(coupled) transport equations or \emph{linear response} equations
or kinetic equations
and the coefficients $L_{a,b}$ are
known as \emph{Onsager coefficients}.
Since we are assuming small thermodynamic forces,
the temperature difference $\Delta T=T_1-T_2$ is
small compared to $T_1\approx T_2\approx T$,
so that $X_2=\Delta T /T^2$.

Perhaps we should also stress that we focus on the stationary and steady state situations, where all the forces and responsive currents  are independent of time, on average, apart from fluctuations.

The positivity of the entropy production rate,
\begin{equation}
\dot{\mathscr{S}}=J_1 X_1 + J_2 X_2\ge 0,
\label{eq:sprod}
\end{equation}
implies for the Onsager coefficients that
\begin{equation}
\left\{
\begin{array}{l}
{\displaystyle
L_{11}\ge 0,
}
\\
{\displaystyle
L_{22}\ge 0,
}
\\
{\displaystyle
L_{11}L_{22}
-\frac{1}{4}\,(L_{12}+L_{21})^2 \ge 0.
}
\end{array}
\right.
\label{dots}
\end{equation}

Assuming the property of time-reversal invariance of the
equations of motion, \textcite{onsager} derived fundamental relations, known
as \emph{Onsager reciprocal relations} for the cross coefficients
of the Onsager matrix: $L_{a,b}=L_{b,a}$.
When an external magnetic field
${\bm B}$ is applied to the system, the laws of physics remain
unchanged if time $t$ is replaced by $-t$, provided that simultaneously
the magnetic field ${\bm B}$ is replaced by $-{\bm B}$. In this
case the Onsager-Casimir relations \cite{onsager,casimir} read
\begin{equation}
L_{a,b}({\bm B})=L_{b,a}(-{\bm B}).
\end{equation}
At zero magnetic field, we recover the Onsager reciprocal relations
$L_{a,b}=L_{b,a}$. Note that only the diagonal coefficients
are bound to be even functions of the magnetic field:
$L_{a,a}({\bm B})=L_{a,a}(-{\bm B})$, while in general, for $a\ne b$,
$L_{a,b}({\bm B})\ne L_{a,b}(-{\bm B})$.

The Onsager coefficients are related to the familiar
transport coefficients. In the case of thermoelectricity
we have
\begin{equation}
G=\left(\frac{J_1}{\Delta V}\right)_{\Delta T=0}=\frac{L_{11}}{T},
\label{eq:el_conductance}
\end{equation}
\begin{equation}
\Xi=\left(\frac{J_2}{\Delta T}\right)_{J_1=0}=
\frac{1}{T^2}\frac{\det {\bm L}}{L_{11}},
\label{eq:th_conductance}
\end{equation}
\begin{equation}
S=-\left(\frac{\Delta V}{\Delta T}\right)_{J_1=0}=
\frac{1}{T}\frac{L_{12}}{L_{11}},
\label{eq:seebeck}
\end{equation}
where $G$ is the (isothermal) \emph{electric conductance},
$\Xi$ the \emph{thermal conductance},
$S$ the \emph{thermopower} (or Seebeck coefficient),
and ${\bm L}$ denotes the Onsager matrix with matrix elements 
$L_{a,b}$ ($a,b=1,2$).
The Peltier coefficient
\begin{equation}
\Pi=\left(\frac{J_2}{J_1}\right)_{\Delta T=0}
=\frac{L_{21}}{L_{11}}
\end{equation}
is related to the thermopower via the Onsager reciprocal 
relation: $ \Pi({\bm B})=TS(-{\bm B})$.
Note that the Onsager-Casimir relations imply 
$G(-{\bm B})=G({\bm B})$ and $\Xi(-{\bm B})=\Xi({\bm B})$,
but in general do not
impose the symmetry of the Seebeck coefficient under the exchange
${\bm B}\to -{\bm B}$.

We can eliminate in the phenomenological equations
(\ref{eq:coupledlinear}) the Onsager matrix elements in favor
of the transport coefficients $G,\Xi,S,\Pi$, thus obtaining
\begin{equation}
\left\{
\begin{array}{l}
J_1=G \Delta V + G S \Delta T,
\\
\\
J_2=G\Pi\Delta V + (\Xi+GS\Pi)\Delta T.
\end{array}
\right.
\end{equation}
By eliminating $\Delta V$ from these two equations we obtain an interesting
interpretation of the Peltier coefficient. Indeed, the entropy current reads
\begin{equation}
J_{\mathscr{S}} = \frac{J_2}{T}=
\frac{\Pi}{T}\,J_1+\frac{\Xi}{T}\Delta T.
\label{eq:entropycurrent}
\end{equation}
Hence, $\Pi/T$ can be understood as the entropy transported 
by the electron flow $J_1$. Since $J_1=e J_\rho$, each 
electron carries an entropy of $e \Pi/T$. 
This contribution to the entropy current adds to the 
last term in (\ref{eq:entropycurrent}), which is independent
of the electric current.
For time-reversal symmetric systems, the same 
interpretation applies to the Seebeck coefficient,
since in this case $S=\Pi/T$.  
The heat flow $J_2=TJ_{\mathscr{S}}$ is the sum of two terms,
$\Pi J_1$ and $\Xi \Delta T$. While the last term is irreversible,
the first one is reversible, that is, it changes sign when 
reversing the direction of the current. 
It can be intuitively understood that efficient energy 
conversion requires to minimize irreversible, dissipative processes 
with respect to reversible processes. Hence, it is desirable to 
have a large Peltier coefficient and a small heat conductance.

The heat dissipation rate $\dot{Q}$ can be 
computed from the entropy production rate 
(\ref{eq:sprod}):
\begin{equation}
\dot{Q}=T \dot{\mathscr{S}}=
\frac{J_1^2}{G}
+\frac{\Xi}{T}(\Delta T)^2
+J_1(\Pi-TS)\frac{\Delta T}{T},
\label{eq:qprod}
\end{equation}
where the first term is the Joule heating, the 
second term is the heat lost by thermal resistance 
and the last term, which disappears for time-reversal 
symmetric systems, can be negative when 
$J_1(\Pi-TS)<0$, thus reducing the dissipated heat. 
It is clear from 
(\ref{eq:qprod}) that to minimize dissipative effects
for a given electric current and thermal gradient, we 
need a large electric conductance and low thermal conductance.

Under the assumption of \emph{local equilibrium},
we can write
coupled equations like (\ref{eq:coupledlinear}),
connecting local fluxes to local forces, expressed
in terms of gradients $\nabla \mu$, $\nabla T$ rather
than $\Delta \mu$, $\Delta T$ (see, for instance,
\textcite{callen}). In this case,
Eqs.~(\ref{eq:el_conductance}) and (\ref{eq:th_conductance})
can be written with on the left-hand side the
\emph{electric conductivity} $\sigma$ and
the \emph{thermal conductivity} $\kappa$ rather than
the conductances $G$ and $\Xi$.


\section{THERMODYNAMIC EFFICIENCIES}

\label{sec:TE}

\subsection{Finite-time thermodynamics}
\label{sec:CA}

A cornerstone result goes back to \textcite{carnot}.
In a cycle between two reservoirs at temperatures $T_1$ and $T_2$ $(T_1>T_2)$,
the efficiency $\eta$, defined as the ratio  of the performed work
$W$ over the heat $Q_1$ extracted from the high temperature reservoir,
is bounded by the \emph{Carnot efficiency} $\eta_C$:
\begin{equation}
\eta = \frac{W}{Q_1} \leq \eta_C = 1-\frac{T_2}{T_1}.
\end{equation}
The Carnot efficiency is obtained for a quasi-static transformation
which requires infinite time and therefore the extracted power, in this
limit, reduces to zero. An important question is how much 
the efficiency deteriorates when the cycle is operated in a 
finite time. This is the central question in the field 
of \emph{finite-time thermodynamics} 
(for a recent review, see \textcite{andresen11}).
In particular, \emph{endoreversible thermodynamics} 
\cite{rubin79,hoffmann97} 
views a thermodynamic system as a collection of reversible 
subsystems which interact in an irreversible manner.

\begin{figure}
\begin{center}
\epsfxsize=75mm\epsffile{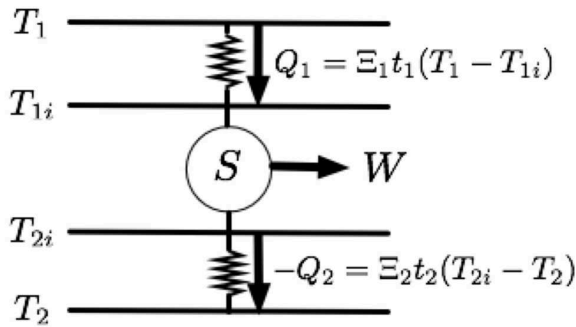}
\caption{Schematic drawing of the endoreversible engine for
the Curzon-Ahlborn cycle. The two heat baths at temperatures
$T_1$ and $T_2$ are coupled for times $t_1$ and $t_2$ 
to the system $S$ (the working fluid, with output work 
per cycle equal to $W$) by heat conductances 
$\Xi_1$ and $\Xi_2$. The system $S$ is considered as a Carnot 
engine operating between the internal temperatures 
$T_{1i}$ and $T_{2i}$ ($T_1>T_{1i}>T_{2i}>T_2$).}
\label{fig:curzon}
\end{center}
\end{figure}

A very important concept is that of \emph{efficiency at
maximum power}. An upper bound for 
the output power $W$ of a heat engine
can be deduced for the endoreversible Curzon-Ahlborn (CA) engine depicted in 
Fig.~\ref{fig:curzon}.
The CA engine consists of two heat baths at temperatures $T_1$ and 
$T_2$ and a reversible Carnot engine operating between internal temperatures
$T_{1i}$ and $T_{2i}$ ($T_1>T_{1i}>T_{2i}>T_2$). 
The two processes of heat transfer, from the 
hot reservoir to the system and from the system to the cold 
reservoir, are the only irreversible processes in the CA engine. 
The output work $W$ is the difference between the heat $Q_1$ absorbed
from the hot reservoir and the heat $-Q_2$ ($Q_1>0$, $Q_2<0$) 
evacuated to the cold 
reservoir ($W=Q_1+Q_2$). 
Heat transfers take place during the isothermal strokes
of the Carnot cycle, with the working fluid (the system) at 
internal temperatures
$T_{1i}$ and $T_{2i}$. We further assume that 
the rate of heat flow $\dot{Q}_1$ ($\dot{Q}_2$) is 
proportional to the temperature difference 
$T_1-T_{1i}$ ($T_2-T_{2i}$) between the hot (cold) reservoir
and the working fluid, the proportionality factor being   
the heat conductance $\Xi_1$ ($\Xi_2$). Therefore, we need a time $t_1$
($t_2$) to transfer an amount $Q_1$ ($Q_2$) of heat, so that
\begin{equation}
Q_1= \Xi_1 t_1 (T_1-T_{1i}),
\end{equation}
\begin{equation}
-Q_2= \Xi_2 t_2 (T_{2i}-T_2).
\end{equation}
Finally, we assume that the time spent in the adiabatic strokes of 
the Carnot cycle is negligible compared to the times of the 
isothermal strokes, so that the total time of cycle is approximately
given by $t=t_1+t_2$. Such assumption is justified if the 
relaxation time for the working fluid is fast enough to allow
to operate quickly the adiabatic trasformations. 
The output power reads
\begin{equation}
P=\frac{W}{t}=\frac{Q_1+Q_2}{t}=\frac{k_1t_1(T_1-T_{1i})+
k_2t_2(T_2-T_{2i})}{t_1+t_2}.
\end{equation} 
Taking into account that the internal Carnot engine operating between
temperatures $T_{1i}$ and $T_{2i}$ has efficiency 
$\eta_{Ci}=1-T_{2i}/T_{1i}=1+Q_2/Q_1$ and using the 
relations $Q_1+Q_2=W$ and $t_j=Q_j/[\Xi_j(T_j-T_{ji})]$,
$(j=1,2)$, we can express the power as
\begin{equation}
P=\frac{\Xi_1\Xi_2 \alpha\beta(T_1-T_2-\alpha-\beta)}{
\Xi_1 \alpha T_2 + \Xi_2 \beta T_1 +\alpha\beta(\Xi_1-\Xi_2)},
\end{equation}
where we have defined $\alpha=T_1-T_{1i}$ and
$\beta=T_{2i}-T_2$. By maximizing the power with respect to 
the internal temperatures $T_{1i}$ and $T_{2i}$ we obtain
\begin{equation}
P_{\rm max}=\Xi_1\Xi_2\left(
\frac{\sqrt{T_1}-\sqrt{T_2}}{\sqrt{\Xi_1}+\sqrt{\Xi_2}}\right)^2.
\end{equation}
The efficiency at the maximum power $P_{\rm max}$,
commonly referred to
as Curzon-Ahlborn upper bound
\cite{Yvon,chambadal,novikov,curzon}, is given by
\begin{equation}
\eta_{CA} = 1-\sqrt{\frac{T_2}{T_1}}=1-\sqrt{1-\eta_C}. \label{eq:ca.eff}
\end{equation}
Remarkably, the CA efficiency is independent of the conductances
$\Xi_1$ and $\Xi_2$.  

It is interesting to remark that the Curzon-Ahlborn efficiency is
invariant under concatenation \cite{vandenbroeck2005}. 
We consider two thermal machines working in a tandem,
the first one between the hot source at temperature 
$T_1$ and a second heat bath at intermediate temperature $T_i$, the second
one between this latter bath and the cold source at temperature $T_2$. 
The first machine absorbs heat $Q_1$, delivers work 
$W'$ and evacuates heat $|Q_i|=Q_1-W'$, the second machine reuses 
heat $|Q_i|$ and outputs work $W''$.
If both machines function at the CA efficiency, also the 
overall efficiency $(W'+W'')/Q_1$ is given by 
$\eta_{CA}=1-\sqrt{T_2/T_1}$. This self-concatenation property
also holds for machines working at Carnot efficiency.

The CA efficiency is not a universal upper bound.
Efficiencies at maximum power not only below, but also above
$\eta_{CA}$ have been reported \cite{mahler08,izumida_okuda08,ss08,elb09,sb2012}.
Yet $\eta_{CA}$ describes the efficiency of
actual thermal plants reasonably well \cite{curzon,esposito2010}, 
and therefore the range of validity of
$\eta_{CA}$ as upper bound 
for the efficiency at maximum power has been widely discussed
in several papers
\cite{vandenbroeck2005,esposito2009,schulman,esposito2010,
linke2010,seifert2011,goupil2012a,goupil2012b,shakouri2012};
for a recent review, see \textcite{tu12}.
The CA bound (\ref{eq:ca.eff}) is an exact and universal bound only 
for systems with (i) time-reversal symmetry and (ii) within a regime of linear
response (see Sec.~\ref{sec:ZT} below).
In the presence of left-right
symmetry in the system, 
the CA bound is exact up to quadratic order in the deviation 
from equilibrium
\cite{esposito2009}.
That is, to second order in $\eta_C$,
\begin{equation}
\eta_{CA}=
\frac{\eta_C}{2}+\frac{\eta_C^2}{8}.
\end{equation} 

The CA efficiency was derived for the Carnot cycle in the limit
of low and symmetric dissipation by \textcite{esposito2010}. 
They considered a Carnot engine which operates under reversible
conditions at the Carnot efficiency when the cycle duration 
becomes infinitely long. In that limit, the system entropy 
increase $\Delta \mathscr{S}=Q_1/T_1$
during the isothermal transformation at the hot temperature
$T_1$ is equal to the system entropy decrease 
$-\Delta \mathscr{S}=Q_2/T_2$ during
the isothermal transformation at the cold temperature $T_2$. 
Hence, there is no overall entropy production 
and the Carnot efficiency $\eta_C=1+Q_2/Q_1=1-T_2/T_1$ is achieved.
\textcite{esposito2010} consider the weak dissipation regime and 
assume that the system relaxation is much faster than the times 
$t_1$ and $t_2$ spent in the isothermal strokes, so that the 
overall cycle duration is to a good approximation given by $t_1+t_2$.  
In the low dissipation regime the entropy production is proportional
to $1/t_1$ and $1/t_2$, so that it vanishes in the limit of 
infinite-time cycle where it is supposed that the Carnot efficiency
is recovered. Therefore the amount of heat entering the system 
from the hot (cold) reservoir is, to first order in $1/t_1$ and
$1/t_2$,
\begin{equation}
Q_1=T_1\left(\Delta \mathscr{S}-\frac{\Sigma_1}{t_1}\right),
\quad
Q_2=T_2\left(-\Delta \mathscr{S}-\frac{\Sigma_2}{t_2}\right),
\end{equation}
with $\Sigma_1$ and $\Sigma_2$ coefficients depending on the 
specific implementation. 
The maximum of the output power 
\begin{equation}
P=
\frac{Q_1+Q_2}{t_1+t_2}=
\frac{(T_1-T_2)\Delta \mathscr{S} - T_1\Sigma_1/t_1-
T_2\Sigma_2/t_2}{t_1+t_2}
\end{equation}
is obtained when $\partial P/\partial t_1=\partial P/\partial t_2=0$.
This leads to the efficiency at the maximum output power
\begin{equation}
\eta(P_{\rm max})=
\frac{\eta_C\left(1+\sqrt{\frac{T_2\Sigma_2}{T_1\Sigma_1}}\right)}{\left(1
+\sqrt{\frac{T_2\Sigma_2}{T_1\Sigma_1}}\right)^2+\frac{T_2}{T_1}
\left(1-\frac{\Sigma_2}{\Sigma_1}\right)}.
\label{eq:etas1s2}
\end{equation}
Note that this result was also obtained in the context 
of stochastic thermodynamics by \textcite{ss08}.
Within linear response,
the Curzon-Ahlborn efficiency is recovered for symmetric dissipation,
$\Sigma_1=\Sigma_2$. From (\ref{eq:etas1s2}) we obtain
\begin{equation}
\eta_-=\frac{\eta_C}{2}\le \eta(P_{\rm max}) \le 
\eta_+=\frac{\eta_C}{2-\eta_C}.
\label{eq:uppermax}
\end{equation}
with the lower and upper bound reached in the limits of 
completely asymmetric dissipation, for  
$\Sigma_2/\Sigma_1\to \infty$ and 
$\Sigma_2/\Sigma_1\to 0$, respectively.
The lower and upper bound coincide in the linear response
regime where $\eta_-=\eta_+=\eta_{CA}=\eta_C/2$.
The same upper bound as in (\ref{eq:uppermax}) was obtained with 
a different approach by \textcite{schulman}.

\subsection{Figure of merit for thermodynamic efficiency}
\label{sec:ZT}

\begin{figure}
\begin{center}
\epsfxsize=80mm\epsffile{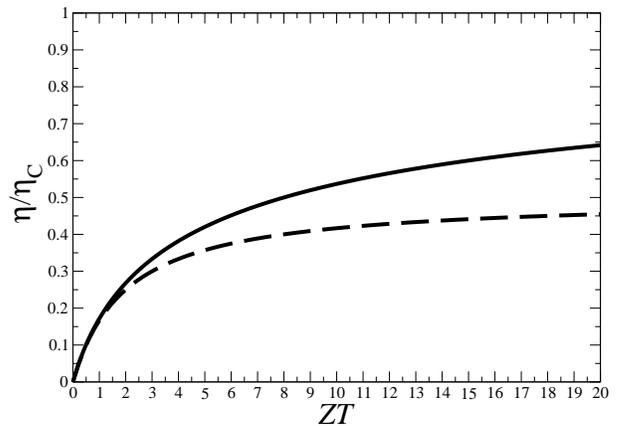}
\caption{Linear response efficiency for heat to work conversion, 
in units of Carnot efficiency $\eta_C$, as a function of the figure
of merit $ZT$. The top and the bottom curve correspond to the 
maximum efficiency $\eta_{\rm max}$ and to the efficiency at 
the maximum power $\eta(P_{\rm max})$, respectively.}
\label{fig:ZT}
\end{center}
\end{figure}

Within the linear response,
the efficiency of steady state heat to work conversion 
reads
\begin{equation}
\eta=\frac{\dot{W}}{\dot{Q}_1}=\frac{-TX_1J_1}{J_2}
=\frac{-TX_1(L_{11}X_1+L_{12}X_2)}{L_{21}X_1+L_{22}X_2},
\label{eq:efficiency}
\end{equation}
where $J_2=\dot{Q}_1>0$ and the power $P=\dot{W}>0$. 
The maximum of $\eta$ over $X_1$, for fixed $X_2$, is achieved
for 
\begin{equation}
X_1=\frac{L_{22}}{L_{21}}\left( 
-1+\sqrt{\frac{\det{\bm L}}{L_{11}L_{22}}}\right)X_2.
\end{equation}
For systems with time-reversal symmetry 
(so that $L_{12}=L_{21}$),
the \emph{maximum efficiency} is given by
\begin{equation}
\eta_{\rm max}=
\eta_C\,
\frac{\sqrt{ZT+1}-1}{\sqrt{ZT+1}+1},
\label{etamaxB0}
\end{equation}
where the \emph{figure of merit}
\begin{equation}
ZT=\frac{L_{12}^2}{\det {\bm L}}
\label{eq:ZT}
\end{equation}
is a dimensionless parameter. In the case of thermoelectricity,
$ZT$, expressed in terms 
of the electric conductance $G$, the thermal conductance
$\Xi$ and the thermopower $S$, reads 
\begin{equation}
ZT=\frac{G S^2}{\Xi}\,T.
\label{eq:ZTthermo}
\end{equation}
For systems with local equilibrium, the figure of merit
can be expressed in therms of the material constants, the electric conductivity $\sigma$ and
the thermal conductivity $\kappa$, rather than $G$ and $\Xi$:
$ZT=(\sigma S^2/\kappa)T$.
The only restriction imposed by thermodynamics 
(more precisely, by the positivity of the entropy production rate) 
is $ZT\ge 0$ and
$\eta_{\rm max}$ is a monotonous growing function 
of $ZT$, with $\eta_{\rm max}=0$ when $ZT=0$ and 
$\eta_{\rm max}\to \eta_C$ when $ZT\to\infty$ (full curve in 
Fig.~\ref{fig:ZT}). 

Note that $ZT$ diverges (thus leading to Carnot efficiency) 
if and only if the the Onsager matrix ${\bm L}$
is ill-conditioned, namely the condition number
\begin{equation}
{\rm cond}({\bm L})=\frac{[{\rm Tr}({\bm L})]^2}{\det {\bm L}}
\end{equation}
diverges and therefore the system
(\ref{eq:coupledlinear}) becomes singular.
That is,
$J_2=c J_1$, the proportionality factor $c$ being independent
of the values of the applied thermodynamic forces.
In short, within linear response (and without external magnetic fields
or other effects breaking time-reversal symmetry) the Carnot efficiency
is obtained if and only if charge and energy flows are proportional
(\emph{tight coupling} condition, also known as strong coupling 
in the literature).

The output power 
\begin{equation}
P=-TX_1J_1 
=-TX_1(L_{11}X_1+L_{12}X_2)
\end{equation}
is maximal when 
\begin{equation}
X_1=-\frac{L_{12}}{2L_{11}}\,X_2
\label{eq:X1max}
\end{equation}
and is given by 
\begin{equation}
P_{\rm max}=\frac{T}{4}\frac{L_{12}^2}{L_{11}}\,X_2^2=
\frac{\eta_C}{4}\,\frac{L_{12}^2}{L_{11}}\,X_2.
\end{equation}
Using Eqs.~(\ref{eq:el_conductance}) and (\ref{eq:seebeck})
we can also write
\begin{equation}
P_{\rm max}=\frac{1}{4}\,S^2 G (\Delta T)^2.
\end{equation}
We can see from this last equation that the maximum power
is directly set by the combination $S^2 G$, known 
for this reason as \emph{power factor}. 
Note that $P$ is a quadratic function of $X_1$ and the
maximum is obtained for the value (\ref{eq:X1max}) 
corresponding to half of the stopping force,
\begin{equation}
X_1^{\rm stop}=-\frac{L_{12}}{L_{11}}\,X_2,
\end{equation}
that is, to the value for which the motion halts,
$J_1=0$.
For systems with time reversal symmetry, 
the efficiency at maximum power 
reads \cite{vandenbroeck2005}
\begin{equation}
\eta(P_{\rm max})=\frac{\eta_C}{2}\frac{ZT}{ZT+2}.
\label{etawmaxB0}
\end{equation}
This quantity also is a monotonous growing function of 
$ZT$, with $\eta(P_{\rm max})=0$ when $ZT=0$ and
$\eta(P_{\rm max})\to \eta_C/2$ when $ZT\to\infty$ (dashed curve in
Fig.~\ref{fig:ZT}). Note that for small $ZT$ we
have $\eta(P_{\rm max})\approx\eta_{\rm max}\approx 
(\eta_C/4) ZT$. The difference between 
$\eta(P_{\rm max})$ and $\eta_{\rm max}$ becomes 
relevant only for $ZT>1$. 

\begin{figure}
\begin{center}
\epsfxsize=80mm\epsffile{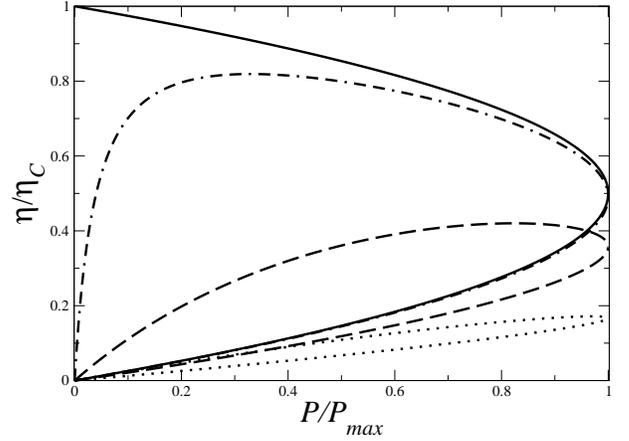}
\caption{Relative efficiency $\eta/\eta_C$ versus normalized 
power $P/P_{\rm max}$. From bottom to top: $ZT=1,5,100$, and $\infty$.
In each curve the lower branch corresponds to a force ratio
$r\le 1/2$, the upper branch to $r\ge 1/2$. Maximum efficiency is 
always achieved on the upper branch.}
\label{fig:eta_power}
\end{center}
\end{figure}

Note that we can establish an efficiency versus power plot. 
We can express the ratio between the power at a given value of
$X_1$ and the maximum power as a function of 
the force ratio $r=X_1/X_1^{\rm stop}$:
\begin{equation}
\frac{P}{P_{\rm max}}=4 r (1-r). 
\end{equation}
This relation can be inverted:
\begin{equation}
r=\frac{1}{2}\,\left[1\pm \sqrt{1-\frac{P}{P_{\rm max}}}\right],
\end{equation}
with the plus sign for $r\ge 1/2$ and the minus sign for $r\le 1/2$.
Inserting this latter relation into Eq.~(\ref{eq:efficiency}) we can 
express the efficiency (normalized to the Carnot efficiency) as 
\begin{equation}
\frac{\eta}{\eta_C}=
\frac{\displaystyle{\frac{P}{P_{\rm max}}}}{
\displaystyle{2\left(1+\frac{2}{ZT}\mp \sqrt{
1-\frac{P}{P_{\rm max}}}\right)}},
\end{equation}
where the minus sign corresponds to $r\ge 1/2$, the plus sign to
$r\le 1/2$.
Plots of the normalized efficiency versus the normalized power
are shown in Fig.~\ref{fig:eta_power}, for several values 
of the figure of merit $ZT$. 
Note that, while for low values of $ZT$ the maximum efficiency 
is close to the efficiency at maximum power, for large $ZT$
the difference becomes relevant (see also Fig.~\ref{fig:ZT}).
For $ZT=\infty$ the Carnot efficiency is achieved at 
the stopping power $X_1=X_1^{\rm stop}$ ($r=1$). 

When the force ratio exceeds one, $r>1$, the thermoelectric device works as 
a \emph{refrigerator}. In this case the most important benchmark is 
the \emph{coefficient of performance} 
$\eta^{(r)}=J_2/P$ ($J_2<0$, $P<0$), given by the ratio of the heat 
current extracted from the cold system over the absorbed power. 
By optimizing this quantity within linear response, we obtain 
\begin{equation}
\eta_{\rm max}^{(r)}=
\eta_C^{(r)}\,
\frac{\sqrt{ZT+1}-1}{\sqrt{ZT+1}+1},
\label{etamaxref}
\end{equation}
where $\eta_C^{(r)}=T_2/(T_1-T_2)\approx 1/(T X_2)$ is the efficiency
of an ideal, dissipationless refrigerator. 
Since the ratio $\eta_{\rm max}^{(r)}/\eta_C^{(r)}$ for refrigeration is equal
to the ratio $\eta_{\rm max}/\eta_C$ for thermoelectric power generation,
$ZT$ is the figure of merit for both regimes. 

\subsection{Systems with broken time-reversal symmetry}

\begin{figure}
\begin{center}
\epsfxsize=80mm\epsffile{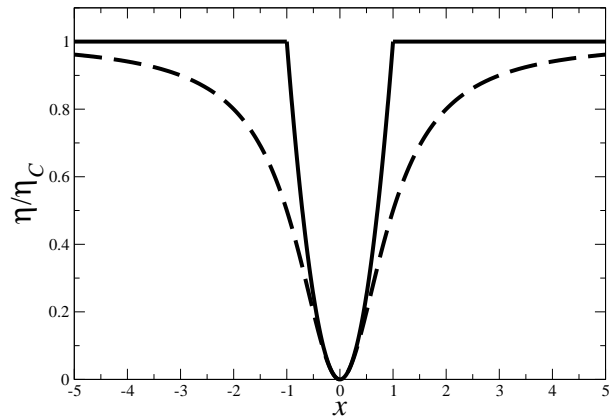}
\caption{Ratio $\eta/\eta_C$ as a function of the asymmetry parameter $x$,
with $\eta=\bar{\eta}(P_{\rm max})$ (dashed curve) and 
$\eta=\bar{\eta}_{\rm max}$
(full curve). For $x=1$, 
$\bar{\eta}(P_{\rm max})=\eta_C/2$ and $\bar{\eta}_{\rm max}=\eta_C$ 
are obtained for $y(x=1)=ZT=\infty$.}
\label{fig:magnetic}
\end{center}
\end{figure}

The same analysis as above can be repeated 
when time-reversal symmetry is broken, say by a magnetic
field ${\bm B}$ (or by other effects 
such as the Coriolis force). In this case the maximum efficiency and the
efficiency at maximum power are both determined
by two parameters \cite{BSC2011}: the asymmetry parameter
\begin{equation}
x=\frac{L_{12}}{L_{21}}=\frac{S({\bm B})}{S(-{\bm B})}
\label{def:x}
\end{equation}
and the ``figure of merit''
\begin{equation}
y=\frac{L_{12}L_{21}}{\det {\bm L}}=
\frac{G({\bm B}) S({\bm B})S(-{\bm B})}{\Xi({\bm B})}\,T.
\end{equation}
The maximum efficiency reads
\begin{equation}
\eta_{\rm max}= \eta_C\,x\,
\frac{\sqrt{y+1}-1}{\sqrt{y+1}+1},
\label{eq:ZTx}
\end{equation}
while the efficiency at maximum power is
\begin{equation}
\eta(P_{\rm max})=
\frac{\eta_C}{2}\,\frac{xy}{2+y}.
\label{etawmax}
\end{equation}
In the particular case $x=1$, $y$ reduces to the $ZT$
figure of merit of the
time-symmetric case, Eq.~(\ref{eq:ZTx}) reduces to
Eq.~(\ref{etamaxB0}),
and Eq.~(\ref{etawmax}) to Eq.~(\ref{etawmaxB0}).
While thermodynamics does not impose any restriction on the
attainable values of the asymmetry parameter $x$, the positivity 
of entropy production implies $h(x)\le y \le 0$ if $x\le 0$ and
$0\le y \le h(x)$ if $x\ge 0$, where the function $h(x)= 4x/(x-1)^2$.
Note that $\lim_{x\to 1} h(x)=\infty$ and therefore there is no 
upper bound on $y(x=1)=ZT$. For a given value of the asymmetry $x$, 
the maximum (over $y$) 
$\bar{\eta}(P_{\rm max})$ of $\eta(P_{\rm max})$ and the maximum
$\bar{\eta}_{\rm max}$ of $\eta_{\rm max}$ are obtained for $y=h(x)$:
\begin{equation}
\bar{\eta}(P_{\rm max})=\eta_C\frac{x^2}{x^2+1},
\label{eq:boundetapmax}
\end{equation}
\begin{equation}
\bar{\eta}_{\rm max}=
\left\{
\begin{array}{ll}
\eta_C\,x^2 & {\rm if}\,\, |x| \le 1,
\\
\\
\eta_C & {\rm if}\,\, |x| \ge 1.
\end{array}
\right.
\label{eq:boundetamax}
\end{equation}
The functions $\bar{\eta}(P_{\rm max})(x)$ and 
$\bar{\eta}_{\rm max}(x)$ 
are drawn 
in Fig.~\ref{fig:magnetic}.
In the case $|x|>1$, it is in principle possible to overcome
the CA limit within linear response and to reach the
Carnot efficiency, for increasingly smaller and smaller figure of merit $y$ as
the asymmetry parameter $x$ increases. The Carnot efficiency is 
obtained for ${\rm det} {\bm L}=(L_{12}-L_{21})^2/4>0$ when $|x|>1$, 
that is, the tight coupling condition is not fulfilled. 

The output power at maximum efficiency reads
\begin{equation}
P(\bar{\eta}_{\rm max})=\frac{\bar{\eta}_{\rm max}}{4}\frac{|L_{12}^2-L_{21}^2|}{L_{11}}\,X_2.
\end{equation}
Therefore, always within linear response,
it is allowed from thermodynamics 
to have Carnot efficiency and nonzero power
simultaneously when $|x|>1$. 
Such a possibility can be understood on the basis of the 
following argument \cite{BSS2013,BS2013}. We first split each current $J_i$
($i=1,2$) into a 
reversible and an irreversible part, defined by
\begin{equation}
J_i^{\rm rev}=\sum_{j=1}^2 \frac{L_{ij}-L_{ji}}{2}\,X_j,
\;\;
J_i^{\rm irr}=
\sum_{j=1}^2 \frac{L_{ij}+L_{ji}}{2}\,X_j.
\label{eq:Jrevirr}
\end{equation}
It is readily seen from Eq.~(\ref{eq:sprod}) and
(\ref{eq:Jrevirr}) that only the irreversible part
of the currents contributes to the entropy production:
\begin{equation}
\dot{\mathscr{S}}=J_1^{\rm irr} X_1 + J_2^{\rm irr} X_2.
\end{equation}
The reversible currents $J_i^{\rm rev}$ vanish for 
${\bm B}=0$. On the other hand, for broken time-reversal symmetry 
the reversible currents can in principle become arbitrarily large, 
giving rise to the possibility of dissipationless transport.

While in the time-reversal case the linear response
normalized maximum efficiency
$\eta_{\rm max}/\eta_C$
and coefficient of performance
$\eta_{\rm max}^{(r)}/\eta_C^{(r)}$ 
for power generation and refrigeration
coincide, this is no longer the case with broken
time-reversal symmetry. For refrigeration 
the maximum value of the coefficient of performance reads
\begin{equation}
\eta_{\rm max}^{(r)}=\eta_C^{(r)} 
\,\frac{1}{x}\,\frac{\sqrt{y+1}-1}{\sqrt{y+1}+1}.
\label{eq:etarefrigeration}
\end{equation}
For small fields, $x$ is in general
a linear function of the magnetic field,
while $y$ is by construction an even function of the field.
As a consequence, a small external magnetic field either improves
power generation and
worsens refrigeration or vice-versa, while the average
efficiency
\begin{equation}
\frac{1}{2}\left[\frac{\eta_{\rm max}({\bm B})}{\eta_C}+
\frac{\eta_{\rm max}^{(r)}({\bm B})}{\eta_C^{(r)}}\right]
=\frac{\eta_{\rm max}({\bm 0})}{\eta_C}=
\frac{\eta_{\rm max}^{(r)}({\bm 0})}{\eta_C^{(r)}},
\end{equation}
up to second order corrections.
Due to the Onsager-Casimir relations, $x(-{\bm B})=1/x({\bm B})$
and therefore by inverting the direction of the magnetic field
one can improve either power generation or refrigeration.

Onsager relations do not impose the symmetry 
$x=1$, i.e., we can have $S({\bm B})\ne S(-{\bm B})$.
However, as discussed in Sec.~\ref{sec:probes} below,
in the non-interacting case
$S(-{\bm B})=S({\bm B})$ as a consequence
of the symmetry properties of the scattering matrix \cite{datta}.
On the other hand,
this symmetry
may be violated when electron-phonon and electron-electron
interactions are taken into account.
While the Seebeck coefficient has always been found
to be an even function of the magnetic
field in two-terminal purely metallic mesoscopic
systems \cite{lsb98,gbm99},
measurements for certain orientations of 
a bismuth crystal \cite{wolfe1963},
Andreev interferometer experiments \cite{chandrasekhar}
and recent theoretical studies \cite{jacquod,SBCP2011,sanchez2011}
have shown that systems
in contact with a superconductor or subject to inelastic
scattering can exhibit non-symmetric thermopower,
i.e., $S(-{\bm B})\ne S({\bm B})$.
So far, investigations of various
classical \cite{horvat2012} and quantum \cite{SBCP2011}
dynamical models have shown arbitrarily large values of the
asymmetry $x$, but correspondingly with low efficiency. 
However, efficiency at maximum power beyond the CA limit for
$x>1$ has been recently shown in \textcite{BSS2013,vinitha2013,BS2013}
(see Sec.~\ref{sec:probes} below).

\section{NON-INTERACTING SYSTEMS, LANDAUER-B\"UTTIKER FORMALISM}
\label{sec:landauer}

Exact calculation of thermodynamic efficiencies is possible
for non-interacting models by means of the Landauer-B\"uttiker approach.
This approach describes the coherent flow of electrons through
a channel. All dissipative and phase-breaking processes are limited
to the contacts (reservoirs).
The electric and thermal
currents are expressed in terms of the scattering
(transmission) properties
of the system \cite{datta,imry}:
\begin{equation}
J_1=\frac{e}{h}
\int_{-\infty}^\infty dE \tau(E) [f_1(E)-f_2(E)],
\label{J1landauer}
\end{equation}
\begin{equation}
J_2=\frac{1}{h}
\int_{-\infty}^\infty dE (E-\mu_1) \tau(E) [f_1(E)-f_2(E)].
\label{J2landauer}
\end{equation}
Here, $e$ is the electron charge, $h$  the Planck's constant,
$\tau(E)$ the \emph{transmission probability} for a particle
with energy $E$ to transit from terminal
(reservoir) 1 to terminal 2
($0\le \tau(E)\le 1$),
and $f_i(E)=\{\exp[(E-\mu_i)/k_B T_i]+1\}^{-1}$ is
the Fermi distribution of the particles
injected from reservoir $i$.
Note that $J_2=\dot{Q}_1$ is the heat
current from the hot reservoir ($T_1>T_2$).

The Onsager coefficients $L_{a,b}$ can be derived from the
linear expansion of the currents
(\ref{J1landauer}) and (\ref{J2landauer}). We obtain
\begin{equation}
L_{11}=e^2 T I_0,\;\;
L_{12}=L_{21}=e T I_1,\;\;
L_{22}=T I_2.
\end{equation}
Here, the integrals $I_n$ have been defined as
\begin{equation}
I_n=\frac{1}{h}\int_{-\infty}^\infty
dE (E-\mu)^n \tau(E) \left(-\frac{\partial f}{\partial E}\right),
\label{eq:In}
\end{equation}
where the derivative of the Fermi function,
$-\partial f/\partial E=1/4k_BT\cosh^2[(E-\mu)/k_BT]$,
is a bell-shaped function centered at $\mu$ and has a width
of the order of $k_B T$.
It immediately follows that \emph{conductances} and
the thermopower can be expressed in terms of the integrals $I_n$:
\begin{equation}
G=e^2 I_0,\;\;
\Xi=\frac{1}{T}\left(I_2-\frac{I_1^2}{I_0}\right),\;\;
S=\frac{1}{eT}\frac{I_1}{I_0}.
\label{eq:landauer.ex}
\end{equation}

While Landauer-B\"uttiker approach describes coherent quantum
transport, semiclassical transport can be described by means
of the Boltzmann equation. Here we consider transport processes that
occur much slower than the relaxation to local equilibrium
and treat collisions within the relaxation-time
approximation \cite{ashcroftmermin}. That is, collisions drive 
the electronic system to local thermodynamic equilibrium under the
assumption that the distribution of electrons emerging from 
collisions does not depend on the structure of their non-equilibrium 
distribution prior to the collision and that collisions do not
alter local equilibrium.  We can then express
\emph{conductivities} and the thermopower in terms of the
integrals
\begin{equation}
K_n=\int_{-\infty}^\infty
dE (E-\mu)^n \Sigma(E) \left(-\frac{\partial f}{\partial E}\right).
\label{eq:Kn}
\end{equation}
Here $\Sigma(E)\approx D(E) t_R(E)\nu(E)^2$ is the transport distribution function, where $D(E)$ is the density of states, $t_R(E)$ the electron
relaxation time, and $\nu(E)$ the electron group velocity.
We obtain \cite{mahansofo}
\begin{equation}
\sigma = e^2 K_0,\;
\kappa=\frac{1}{T}\left(K_2-\frac{K_1^2}{K_0}\right),\;
S=\frac{1}{eT}\frac{K_1}{K_0}.
\label{eq:boltzmann.ex}
\end{equation}

\subsection{Energy filtering}
\label{sec:energyfiltering}

An interesting question is what transmission function
$\tau(E)$ (or transport distribution function $\Sigma(E)$ in
the Boltzmann approach) provides the largest
thermodynamic efficiencies.
\textcite{mahansofo}
showed that, within linear response,
a delta-shaped transmission function leads
to Carnot efficiency.
As discussed in Sec.~\ref{sec:ZT}, 
$ZT$ diverges if and only if the Onsager matrix ${\bm L}$
is ill-conditioned, that is, in the tight coupling limit
$J_2=c J_1$, with $c$ independent of the applied forces.
The tight coupling condition is achieved when the transmission
is possible only within a tiny energy window around $E=E_\star$
(\emph{energy filtering}).
In this case from Eq.(\ref{eq:In}) we obtain
$I_n\approx (E_\star-\mu)^n I_0$, and therefore
\begin{equation}
ZT=\frac{GS^2}{\Xi}\,T=\frac{I_1^2}{I_0I_2-I_1^2}\to\infty.
\end{equation}

The energy filtering mechanism allows us to achieve the Carnot
efficiency also beyond linear response \cite{linke2002,linke2005}.
In this case, assuming $T_1>T_2$, $\mu_1<\mu_2$, $J_1> 0$
and $J_2> 0$, the efficiency for power generation
is given by
\begin{eqnarray}
\begin{array}{c}
{\displaystyle
\eta=\frac{[(\mu_2-\mu_1)/e] J_1}{J_2}
}
\\
\\
{\displaystyle
=
\frac{(\mu_2-\mu_1)\int_{-\infty}^\infty dE \tau(E)
[f_1(E)-f_2(E)]}{\int_{-\infty}^\infty
dE (E-\mu_1) \tau(E) [f_1(E)-f_2(E)]}.
}
\end{array}
\end{eqnarray}
When the transmission is possible only within a tiny energy
window around $E=E_\star$, the efficiency reads
\begin{equation}
\eta= \frac{\mu_2-\mu_1}{E_\star-\mu_1}.
\label{eq:etalinke}
\end{equation}
We have $f_1(E_\star)=f_2(E_\star)$,
namely the occupation of states is the same in the two reservoirs
at different temperatures and electrochemical potentials,
when
\begin{equation}
\frac{E_\star-\mu_1}{T_1}=
\frac{E_\star-\mu_2}{T_2}\Rightarrow
E_\star=\frac{\mu_2T1-\mu_1T_2}{T_1-T_2}.
\label{eq:etalinke2}
\end{equation}
Substituting such $E_\star$ into Eq.~(\ref{eq:etalinke}),
we obtain the Carnot efficiency $\eta=\eta_C=1-T_2/T_1$.
Note that Carnot efficiency is obtained in the limit
$J_1\to 0$, corresponding to \emph{reversible transport}
(zero entropy production) and
\emph{zero output power}.

High values of $ZT$ can still be achieved if rather
than delta-shaped transmission function one considers 
sharply rising transmission functions \cite{shakouri2004,humphrey2005}.
The advantage of step over narrow transmission functions 
is that good efficiencies can be obtained
without greatly reducing power.  

\subsection{Noise and probe reservoirs}
\label{sec:probes}

\begin{figure}
\begin{center}
\epsfxsize=80mm\epsffile{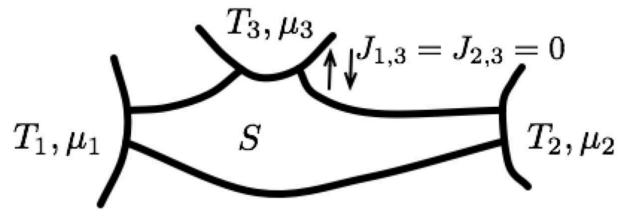}
\caption{Schematic drawing of partially-coherent thermoelectric 
transport, with the third terminal acting as a probe reservoir
mimicking inelastic scattering. The temperature $T_3$ and the chemical
potential $\mu_3$ of the third reservoir are such that the net 
average electric and heat currents through this reservoir vanish:
$J_{1,3}=J_{2,3}=0$. This setup can be generalized to any number
of probe reservoirs, $k=3,...,n$, by setting 
$J_{1,k}=J_{2,k}=0$ for all probes.} 
\label{fig:3ter}
\end{center}
\end{figure}

The Landauer-B\"uttiker approach provides a rigorous framework
for the description of coherent quantum transport.
The transport can be only partially coherent when \emph{inelastic
scattering}, due to the interactions of the electrons with phonons,
photons, and other electrons, is taken into account.
A very convenient way to introduce inelastic scattering is by means
of a third terminal (or \emph{conceptual probe}),
whose parameters (temperature and chemical potential)
are chosen self-consistently so that there is no net \emph{average} flux of
particles and heat between this terminal and the system
(see Fig.~\ref{fig:3ter}).
In mesoscopic physics, probe reservoirs
are commonly used to simulate phase-breaking processes in
partially coherent quantum transport, since they
introduce phase-relaxation without energy damping \cite{buttiker1988}.
The advantage of such approach lies in its simplicity and independence
from microscopic details of inelastic processes.
Probe terminals have been widely used in the literature
and proved to be useful to unveil nontrivial aspects of
phase-breaking
processes \cite{datta}, heat transport and rectification
\cite{visscher,lebowitz,dhar2007,dhar,lebowitz2009,pereira,segal05,segal,saito06},
and thermoelectric transport
\cite{jacquet2009,imry2010,buttiker2011,buttiker2013,buttikerNJP2013,SBCP2011,sanchez2011,imry2012,entin2012,sb2012,buttiker2012,horvat2012,imry2013,imryNJP2013,ruokola2012,muttalib2013,segal2013}.
Note that some of the above models consider the third terminal 
as a bosonic (phonons, photons, or magnons) 
rather than a fermionic bath and cannot be
treated within the Landauer-B\"uttiker approach for non-interacting particles.
In that case one has to use other methods 
such as the Keldysh technique \cite{imry2010,entin2012}. 

The approach can be generalized to any number $n_p$ of probe reservoirs.
We call $J_{1,k}$ and $J_{2,k}$ the charge and heat currents
from the $k$th terminal (at temperature $T_k$ and electrochemical
potential $\mu_k$), with $k=3,...,n$ denoting the 
$n_p=n-2$ probes.
Due to the steady-state constraints of charge and energy conservation,
$\sum_k J_{1,k}=0$ and $\sum_k (J_{2,k}+\mu_k J_{1,k})=0$, 
we can express, for instance, the currents from the 
second reservoir as a function of the remaining $2(n-1)$ currents.
The corresponding generalized forces are given by 
$X_{1,k}=\Delta \mu_k/(eT)$ and $X_{2,k}=\Delta T_k/T^2$, 
with $\Delta \mu_k=\mu_k-\mu$, 
$\Delta T_k= T_k-T$, $\mu=\mu_2$, and $T=T_2$.
The linear response relations between currents and thermodynamic
forces, 
\begin{equation}
J_{i,k}=\sum_{j=1}^2\sum_{\substack{l=1
\\(l\ne 2)}}^n L_{ik;jl} X_{j,l},
\end{equation}
are expressed in terms of an Onsager matrix ${\bm L}$ of 
size $2(n-1)$. 
We then impose the condition of zero average currents through 
the probes, $J_{1,k}=J_{2,k}=0$ for $k=3,...,n$ to reduce 
the Onsager matrix to a $2\times 2$ matrix ${\bm L}'$ 
connecting the fluxes $J_{i,1}$ through the first reservoir
and the conjugated forces $X_{i,1}$:
\begin{equation}
 \left( \begin{array}{c}
  J_{1,1} \\
  J_{2,1}
\end{array}
\right) =  \left( \begin{array}{cc}
              L'_{11} &  L'_{12} \\
              L'_{21} & L'_{22}
            \end{array} \right) \left( \begin{array}{cc}
              X_{1,1} \\
             X_{2,1}
            \end{array} \right).
\label{eq:Lred}
\end{equation}
The reduced matrix ${\bm L}'$ fulfills the Onsager-Casimir
relations and represents the Onsager matrix for two-terminal
inelastic transport modeled by means of self-consistent reservoirs.
Details of the reduction from ${\bm L}$ to ${\bm L}'$ for $n=3$
reservoirs are provided in \textcite{SBCP2011}.
Note that, if the average flow of particles
and heat through the
probe reservoirs is zero, then thermodynamic efficiencies can
be computed by means of the standard two-terminal formulas
(\ref{eq:ZTx}) and (\ref{etawmax}), with the parameters
$x=L'_{12}/L'_{21}$ and $y=L'_{12}L'_{21}/{\rm det} {\bm L}'$.
However, the transport between these two terminals is no longer
fully coherent, since the electrons can be absorbed and emitted
by the probe terminals. 

Electric and
heat current can be conveniently computed, for any number of probes, by
means of the multi-terminal Landauer-B\"uttiker formula \cite{datta}:
\begin{equation}
J_{1,k}=\frac{e}{h}\int_{-\infty}^{\infty}
dE \sum_l[\tau_{l\leftarrow k}(E)f_k(E)-\tau_{k\leftarrow l}(E)f_l(E)],
\end{equation}
\begin{eqnarray}
\begin{array}{c}
{\displaystyle
J_{2,k}=\frac{1}{h}\int_{-\infty}^{\infty}
dE (E-\mu_k) \sum_l[\tau_{l\leftarrow k}(E)f_k(E)
}
\\
\\
{\displaystyle
-
\tau_{k\leftarrow l}(E)f_l(E)],
}
\end{array}
\end{eqnarray}
where $\tau_{l\leftarrow k}(E)$ is the transmission probability
from terminal $k$ to terminal $l$ at the energy $E$.
Charge conservation and the requirement
of zero current at zero bias impose
\begin{equation}
\sum_k\tau_{k\leftarrow l}=
\sum_{k}\tau_{l\leftarrow k}=M_l,
\label{eq:multiterminal}
\end{equation}
with $M_l$ being the number of modes in the lead $l$. Moreover, in 
the presence of a magnetic
field ${\bm B}$ we have
\begin{equation}
\tau_{k\leftarrow l}({\bm B})=\tau_{l\leftarrow k}(-{\bm B}).
\label{eq:taumag}
\end{equation}
The last relation is a consequence of the unitarity of the 
scattering matrix $\mathbb{S}({\bm B})$ that relates the 
outgoing wave amplitudes to the incoming wave amplitudes at the 
different leads. The time reversal invariance of unitary 
dynamics leads to $\mathbb{S}({\bm B})={}^t\mathbb{S}(-{\bm B})$,
which in turn implies (\ref{eq:taumag}) \cite{datta}.
In the two-terminal case, Eq.~(\ref{eq:multiterminal}) means
$\tau_{1\leftarrow 2}=\tau_{2\leftarrow 1}$. Hence, we can
conclude from this relation and Eq.~(\ref{eq:taumag}) that
$\tau_{2\leftarrow 1}({\bm B})=\tau_{2\leftarrow 1}(-{\bm B})$,
thus implying that the Seebeck coefficient is a symmetric
function of the magnetic field. 

The third (probe) terminal
can break the symmetry of the Seebeck coefficient.
We can have $S(-{\bm B})\ne S({\bm B})$, that is, 
$L'_{12}\ne L'_{21}$ in the reduced Onsager matrix
${\bm L}'$. Arbitrarily large values of the asymmetry 
parameter $x=S({\bm B})/S(-{\bm B})=L'_{12}/L'_{21}$
were obtained in \textcite{SBCP2011} by means
of a three-dot Aharonov-Bohm interferometer model.
The asymmetry was found also for 
chaotic cavities,
ballistic microjunctions \cite{sanchez2011}, and
random Hamiltonians drawn from the Gaussian unitary
ensemble \cite{vinitha2013}.
In \textcite{sanchez2011} it was shown tha the asymmetry is
a higher-order effect in the Sommerfeld expansion and therefore 
disappears in the low temperature limit.
The asymmetry was demonstrated also in the framework
of classical physics, for a three-terminal deterministic
railway switch transport model \cite{horvat2012}.
In such model, only the values
zero and one are allowed for the transmission functions
$\tau_{j\leftarrow i}(E)$, i.e., $\tau_{j\leftarrow i}(E)=1$
if particles injected from terminal $i$ with energy $E$ 
go to terminal $j$ and $\tau_{j\leftarrow i}(E)=0$ is such
particles go to a terminal other than $j$. The transmissions
$\tau_{j\leftarrow i}(E)$ are piecewise constant
in the intervals $[E_i,E_{i+1}]$, $(i=1,2,...)$, with switching
$\tau_{j\leftarrow i}=1\to 0$ or viceversa possible at the  
threshold energies $E_i$, with the constraints (\ref{eq:multiterminal}) 
always fulfilled. 

In all the above instances, it was not possible to find at the same
time large values of asymmetry parameter (\ref{def:x}) and high thermoelectric efficiency.  
Such failure was explained by \textcite{BSS2013} and is generic
for non-interacting three-terminal systems. In that case, when 
the magnetic field ${\bm B}\ne 0$, current conservation, which is
mathematically expressed by unitarity of the scattering matrix
$\mathbb{S}$, imposes bounds on the Onsager matrix stronger than 
those derived from positivity of entropy production. We have
\begin{equation}
L_{11}L_{22}
-\frac{1}{4}\,(L_{12}+L_{21})^2 \ge 
\frac{3}{4}\,(L_{12}-L_{21})^2.
\label{eq:3terbound}
\end{equation}
Such constraint reduces to the third inequality of 
Eq.~(\ref{dots}) only in the time-symmetric case 
$L_{12}=L_{21}$, while it is in general a stronger inequality, 
since the right-hand side of Eq.~(\ref{eq:3terbound}) 
is strictly positive when $L_{12}\ne L_{21}$. 
As a consequence, Carnot efficiency can be achieved 
in the three-terminal setup only in the time-symmetric case
${\bm B}=0$. On the other hand, the Curzon-Ahlborn linear response
bound $\eta_{CA}=\eta_C/2$ for the efficiency at maximum power 
can be overcome for moderate asymmetries, $1<x<2$, with a
maximum of $4\eta_C/7$ at $x=4/3$. The bounds obtained by 
\textcite{BSS2013} are in practice saturated in a 
quantum transmission model reminiscent of the 
above described railway switch model \cite{vinitha2013}.
Multi-terminal cases with more than three terminals were also discussed for
noninteracting electronic transport \cite{BS2013}.
By increasing the number $n_p$ of probe terminals, 
the constraint from current conservation on the maximum efficiency
and the efficiency at maximum power becomes weaker than that
imposed by (\ref{eq:3terbound}). However, the bounds 
(\ref{eq:boundetapmax}) and (\ref{eq:boundetamax})
from the second law of
thermodynamics are saturated only in the limit $n_p\to\infty$. 
It is an interesting open question whether 
similar bounds on efficiency, tighter
that those imposed by the positivity of entropy production,
exist in more general transport models for interacting systems.

Probe-reservoir models unveil several other nontrivial aspects of
inelastic processes. For instance, the third reservoir may be a
phonon bath connected to a nanostructure (e.g., a molecule), and it has
been shown that such setup can be very favorable for thermoelectric
energy conversion \cite{imry2012}, notably the setup can act as a 
refrigerator for the local phonon system. 
The \emph{cooling by heating} phenomenon can also be interpreted in 
terms of a third, photonic terminal powering refrigeration. 
\textcite{pekola2007} 
(see also \textcite{pekola2011,pekola2012,vandenbroeck2006})
considered the case in which the photons
emitted by a hot resistor can extract 
heat from a cold metal, providing the energy needed to 
electrons to tunnel to a superconductor (separated from the metal
by a thin insulating junction; no voltage is applied over the
junction). If the temperature of the resistor
is suitably set, only the high energy electrons are removed
from the metal, thus cooling it. Such \emph{Brownian refrigerator}
is still to be experimentally demonstrated.
Similar mechanisms have been discussed for cooling 
a metallic lead, connected to another, higher temperature
lead by means of two adjoining quantum dots \cite{vandenbroeck2012}
or for cooling an optomechanical system \cite{eisert2012}.
In both cases, refrigeration is powered by absorption of photons.

\section{INTERACTING SYSTEMS}
\label{sec:interacting}

\subsection{Green-Kubo formula}
\label{sec:Kubo}

The Green-Kubo formula expresses linear response transport coefficients
in terms of \emph{dynamic correlation functions} of the corresponding
current operators, calculated at thermodynamic equilibrium (see for
instance \textcite{kubo,mahan}):
\begin{eqnarray}
L_{a,b} &=& \lim_{\omega\to 0} {\rm Re} L_{a,b} (\omega), \label{eq:kubo}\\
L_{a,b}(\omega) &=& \lim_{\epsilon\to 0}
\int_0^\infty\!\!\!\!dt e^{-i(\omega\!-\!i\epsilon)t}\lim_{\Omega\to\infty}\frac{1}{\Omega}
\int_0^\beta\!\!\!\!d\tau\langle \hat{J}_a \hat{J}_b (t+i\tau)\rangle, \nonumber
\label{eq:GK}
\end{eqnarray}
where $\beta=1/k_B T$ ($k_B$ it the Boltzmann constant),
$\langle \; \cdot \;\rangle = \left\{{\rm tr}(\;\cdot\;) \exp^{-\beta H}\right\}/{\rm tr} \exp(-\beta H) $
denotes the thermodynamic expectation value, $\Omega$ is the system's volume, and
the currents are $J_a=\langle {\hat J}_a \rangle$, with $\hat{J}_a$ the current operator.
Note that in extended systems, the operator ${\hat J}_a=\int_\Omega d\vec{r} {\hat j}(\vec{r})$ is an extensive quantity,
${\hat j}_a(\vec{r})$ is the current density operator, satisfying the continuity equation
\begin{equation}
\frac{d\hat{\rho}_a(\vec{r},t)}{d t} = \frac{i}{\hbar}\,[H,\hat{\rho}_a] = 
-\nabla \cdot {\hat{j}}(\vec{r},t),
\label{eq:continuity}
\end{equation}
where $\hat{\rho}_a$ is the density of the corresponding conserved quantity, 
say, energy, electric charge, magnetization etc.  
Eq.~(\ref{eq:continuity}) can be equally well written in classical 
mechanics, provided the commutator is substituted by the Poisson
bracket multiplied by the factor $i\hbar$. 
The real part of $L_{a,b}(\omega)$ can be decomposed into a
$\delta$-function at zero frequency defining a generalized \emph{Drude weight}
$D_{a,b}$ (for $a=b$ this is the conventional Drude weight) and a regular part $L_{a,b}^{\rm reg}(\omega)$:
\begin{equation}
{\rm Re} L_{a,b}(\omega)=2\pi D_{a,b}\delta(\omega)+L_{a,b}^{\rm reg}(\omega).
\label{eq:Lreg}
\end{equation}
The matrix of Drude weights can be within linear response also expressed in terms of time-averaged current-current correlations directly:
\begin{equation}
D_{a,b}=\lim_{\bar{t}\to\infty}\frac{1}{{\bar{t}}}\int_0^{\bar{t}} 
dt \lim_{\Omega\to\infty}\frac{1}{\Omega}\int_0^\beta d\tau\langle \hat{J}_a \hat{J}_b (t+i\tau)\rangle.
\label{drudematrix}
\end{equation}
Note that for finite systems, i.e. disregarding the thermodynamic limit $\Omega\to\infty$, it is possible to give a spectral representation of
both $D_{a,b}$ and $L_{a,b}^{\rm reg}$
in terms of the eigenenergies and eigenstates of the
system and of the corresponding Boltzmann weights
\cite{kohn,zotos1997,zotosreview}.

The linear response Kubo formalism has been recently used to
investigate the thermoelectric properties of
one-dimensional integrable and non-integrable
strongly correlated lattice models 
\cite{chaikin1976,furukawa2005,prelovsek2005,shastry2007,shastry2009,shastry2010,georges2013,shastry2013}.
Thermoelectrics of strongly correlated materials are of fundamental 
interest. Moreover, experimental results have 
revealed that some materials, such as sodium cobalt oxide, posses
unusually large thermopower \cite{terasaki97,wang2003}, due in part to strong 
electron interactions \cite{shastry2007b}.

\subsection{Conservation laws and thermoelectric transport}
\label{sec:cmotion}

In interacting (quantum) many-body systems one has to often resort to more abstract mathematical frameworks to describe linear response theory, Green-Kubo formulae and Onsager reciprocity relations \cite{jaksic}.
For describing correlated transport in {\em extended} interacting quantum systems with local interaction, one can use the formalism of $C^*$ algebraic dynamical systems \cite{bratelli}. There we can take advantage of
an {\em emerging causality} \cite{bravyi} as the consequence of the Lieb-Robinson bounds stating that correlations, even in nonrelativistic systems with local interactions, propagate with a finite maximum velocity which is essentially given by the strength of the
interaction. In addition, one can use the Mermin-Wagner theorem to conclude that in one- and two-dimensional systems the static correlations always decay exponentially in non-zero temperature (Gibbsian) equilibrium. Consequently, one can 
prove finite-temperature
ballistic transport 
for systems with \emph{relevant} {\em local} conservation laws \cite{prosen11,ip2012}. 
Similarly, one can prove \cite{prosen2013} 
nonvanishing lower bounds on Green-Kubo diagonal diffusive transport coefficients $L_{a,a}$ (\ref{eq:kubo}) in terms of quadratically-extensive conserved quantities (in non-ballistic cases when linearly-extensive -- local conserved quantities do not exist, or are irrelevant, such
as in the case of one-dimensional half-filled fermi Hubbard chain or non-magnetized Heisenberg spin $1/2$ chain).
A constant of motion $\hat{K}_m$ is by definition
relevant if $\langle \hat{J} \hat{K}_m \rangle \neq 0$, where $J$ 
is the current under consideration.
Ballistic finite-temperature transport as a consequence of the existence 
of relevant conservation laws is a typical feature of 
completely integrable strongly interacting systems,
as suggested some time ago by  \textcite{zotos1997}
(see also \textcite{rosch2001,meisner2005}). 
In a similar way, the theory of quantum integrable systems implies 
ballistic coupled transport, i.e.
thermoelectric and thermomagnetic, properties of such systems, e.g. 
in the anisotropic Heisenberg XXZ spin 1/2 
chain \cite{furukawa2005,sakai}.

More specifically, let us consider a strongly interacting system with a set of $M$ constants of motion $\hat{K}_m$, $m=1,\ldots,M$, i.e. Hermitian operators $\hat{K}_m$ which commute with the Hamiltonian and among themselves, $[H,\hat{K}_m]=0, [\hat{K}_m,\hat{K}_l]=0$, and which
can always be chosen to be orthogonal,
i.e. $\ave{\hat{K}_m \hat{K}_l} = \delta_{k,l}\ave{\hat{K}_m^2}$, via a Gram-Schmidt procedure. Furthermore, let us assume that $\hat{K}_m$ are linearly extensive, i.e. $\ave{\hat{K}_m^2} \propto \Omega$. Then, provided the set $\{ \hat{K}_m\}$ exhausts {\em all} extensive conserved quantities 
(in the thermodynamic limit $\Omega \to \infty$), the matrix of generalized Drude weights (\ref{drudematrix}) can be expressed explicitly by means of Suzuki's formula \cite{suzuki,benenti13}:
\begin{equation}
D_{a,b} = \lim_{\Omega\to\infty}\frac{1}{2\Omega} \sum_{m=1}^M \frac{ \ave{\hat{J}_a \hat{K}_m} \ave{\hat{J}_b \hat{K}_m}}{\ave{\hat{K}_m^2}}.
\label{eq:drude}
\end{equation}
At zero frequency $\omega=0$, the elements of Onsager matrix $L_{a,b}$ can be 
replaced by $D_{a,b}$ in the expression for the figure of merit of thermoelectric, thermomagnetic or thermochemical efficiency $ZT$. 

Conservation laws have an interesting consequence for thermoelectric
efficiency, when there is a single relevant conserved
quantity, $M=1$ in Eq.~(\ref{eq:drude}). In that case
all Onsager matrix elements $L_{a,b}$ are size independent and
therefore also the electric conductance $G\propto L_{11}$ 
and the thermopower $S\propto L_{12}/L_{11}$ are size independent.
On the other hand, the thermal conductance 
$\Xi\propto {\rm det}({\bm L})/L_{11}$ drops with the system size. 
Indeed the Drude weight contribution 
to ${\rm det}({\bm L})$ vanishes, since it is given by 
$D_{11}D_{22}-D_{12}^2$, which vanishes 
as a consequence of Eq.~(\ref{eq:drude}) which for $M=1$ states that the matrix $D_{a,b}$ has rank $1$. 
Since the electric conductance is ballistic, the thermopower   
size-independent and the thermal 
conductance subbalistic (i.e., it drops with the system size), 
we can conclude that the figure of merit $ZT=(GS^2/\Xi)T$ diverges 
with the system size \cite{benenti13}. Note that these 
conclusions for the thermal conductance and the figure of merit
do not hold when $M>1$, as is typical for completely integrable systems.  
In that case we have, in general, 
$D_{11}D_{22}-D_{12}^2\ne 0$, so that thermal conductance 
is ballistic and therefore $ZT$ is size-independent.
This result is not limited to quantum systems and has no dimensional 
restrictions; it has been illustrated by means of a diatomic 
chain of hard-point colliding particles \cite{benenti13},
where the divergence of the figure of merit with the system 
size \cite{casati09} cannot be explained in terms of the energy
filtering mechanism \cite{saito10}, 
and in a two-dimensional 
system connected to reservoirs \cite{monasterio13}, 
with the dynamics simulated by
the multiparticle collision dynamics method \cite{kapral}. 
In both (classical) models collisions are elastic 
and the component of momentum along the direction of 
the charge and heat flows is the only relevant constant of motion. 
Divergence of $ZT$ has been also predicted,
by different theoretical arguments, for a quantum wire 
with weak electron-electron interactions, 
in the limit of infinite wire length \cite{matveev}.

When the underlying many-body system is strongly non-integrable, or when all the extensive local conservation laws are irrelevant for the transporting quantities, i.e. when
$\ave{\hat{J}_a \hat{K}_m}=0$ for all $m$, then the transport is typically diffusive. In the latter case one often finds diffusive transport even if the system is completely integrable \cite{steinigeweg,pz2009}. Then, within the linear response approach, one has
to apply the regularized Onsager matrix $L_{a,b}^{\rm reg}(\omega)$
of Eq.~(\ref{eq:Lreg}) for estimation of $ZT$.

\subsection{Thermalization}
\label{sec:thermalization}

One of the essential prerequisites for using the methods of canonical statistical mechanics is establishing precise conditions under which the system discussed can be claimed to be in thermal equilibrium. 
The understanding of thermalization in closed (isolated) but complex
quantum systems is one of the most
intriguing problems in quantum physics, with deep connections with the
field of quantum chaos \cite{aberg90,deutsch91,srednicki94,izrailev,jacquoddima97,BCS2001}. 
Theoretical interest in these issues 
resurfaced periodically, until very recently \cite{rigol,silva}. 
Recent progress has been made in understanding the conditions under which closedbut complex systems undergo relaxation to equilibrium. 
The conditions for thermalization
are essentially related to the systems' integrability and localization properties (e.g. due to disorder). Non-ergodic systems, such as those possessing some number of exact local conservation laws $\hat{K}_m$ (for example, completely integrable systems) undergo relaxation to a generalized Gibbs state \cite{barthel}, which can again facilitate application of standard statistical mechanics methods.

Thermalization of simple or complex quantum systems immersed into large (many-body), complex, or chaotic environment can be described very conveniently by the methods of open quantum systems \cite{breuer}.  One can treat also extended many-body systems, say particle (or spin) chains, in this way by using a convenient setup in which only boundary (local, on-site) degrees of freedom are directly coupled with the environment. Within the Markovian approximation,
the system's many-body density matrix undergoes a time-evolution dictated by the
Lindblad-Gorini-Kossakowski-Sudarshan equation \cite{lindblad,gorini}
\begin{equation}
\frac{d}{dt}\rho(t) = \hat{\cal L}\rho(t),
\label{eq:lind}
\end{equation}
where the Liouvillian superoperator is defined as
\begin{equation}
\hat{\cal L}\rho := -\frac{i}{\hbar}\,[H,\rho] + \sum_{\mu} \left(L_\mu  \rho L^\dagger_\mu - \frac{1}{2}\,\{L_\mu^\dagger L_\mu,\rho\}\right).
\end{equation}
The Hamiltonian $H$ is here considered to be a sum of locally interacting terms, $H=\sum_{n=1}^{N-1} h_{[n,n+1]}$ and $L_\mu$ are the Lindblad (or so-called quantum jump) operators, which are in the simplest case supported only at the boundary sites, $n=1$ or $n=N$, of the system.
 This setup can be justified \cite{BCPRZ2009} in the regime of weak tunneling interaction between different constituents (particles, spins) of the system, but provides also a more general paradigm of open many-body quantum systems with fully coherent bulk dynamics and incoherent boundary conditions, which is particularly suited for studying non-equilibrium steady state phenomena, such as quantum transport \cite{wichterich2007}. It has been demonstrated 
by numerical simulations that coupling a many-body locally interacting quantum system to a pair of equal Markovian baths through the system's ends in this way, results in thermalization if, and only if, the system's bulk Hamiltonian is {\em not} integrable (via Bethe ansatz) \cite{znidaricetal2010}. Further theoretical work is needed to deeper understand these results.

\subsection{Local equilibrium and non-equilibrium steady states}
\label{sec:local_equilibrium}

Using the just described boundary driven open system's setup, but putting small biases on the rates with which Lindblad jump operators $L_\mu$ target certain locally canonical states near the boundaries, one can address also the problem of bulk transport properties of the system. For example, measuring expectation values of the current observables in the steady states of the Lindblad equation reveals, in the thermodynamic limit $N\to\infty$,  bulk conductivities  \cite{pz2009}, and in principle also the off-diagonal elements of the Onsager matrix. Nevertheless, the problem of establishing local equilibrium in such situations seems to be quite nontrivial \cite{pz2010}. Namely, fixed points of Liouvillean dynamics can describe a variety of qualitatively distinct non-equilibrium quantum phases, ranging from equilibrium-like states where spatial correlations decay exponentially and where local
equilibrium can be well defined, to 
states where spatial correlations only depend on the bias (voltage, temperature drop, etc.) between the reservoirs, but not on the system size. In the latter case one lacks any sensible definition of local equilibrium.
Nevertheless, the approach to non-equilibrium steady states in terms of the methods of open quantum systems appears to be very promising, in particular since strong interactions in the system can easily be treated. It is, however, difficult to control the 
dependence of the results on the type of reservoirs used, as the reservoirs' degrees of freedom are traced out in the very setup.

Markovian master equation models can also be used to treat heat and particle/spin transport in models with conservative noise in the bulk (e.g. dephasing noise which conserves
the number of quasi-particle excitations) \cite{znidaric10,manzano12}. Such models of noise, close in spirit to the probe terminals discussed in
Sec.~\ref{sec:probes}, can be interesting for applications to coupled transport.
They offer elegant ways of treating unwanted degrees of freedom in the bulk, such as lattice vibrations, and they often lead to surprising results, such as noise-induced enhancement of transport \cite{clark13}.

Complementarily, instead of using quantum master equations, one may use another, perhaps more standard approach to non-equilibrium steady states via a Keldysh formalism of non-equlibrium Green's functions  \cite{HaugJauho}, 
where one essentially discusses the scattering of elementary quasi-particle excitations between two or more infinite non-interacting Hamiltonian reservoirs. 
This approach was used, among other things, to study heat transport 
in driven nanoscale engines \cite{arrachea2007,arrachea2012} and 
spin heterostructures \cite{arrachea2009}.
Clearly, for this method to be feasible the full self-energy of the central system has to be somehow tractable. Thus the Keldysh technique 
is usually used when the bulk dynamics is simple
and details of coupling to the reservoirs (i.e. physical leads) are important. 
Open quantum system's approach, on the other hand, has the advantage of allowing numerical and non-perturbative treatments for the many-body central (bulk) Hamiltonians. Another setup in which the Keldysh treatment of infinite Hamiltonian reservoirs can be explicitly implemented is when the bulk dynamics can be treated with integrable effective quantum field theory. For example, if the bulk Hamiltonian is critical, having massless (gapless) low energy excitations and if the temperatures of the reservoirs are small, one may use conformal field theory to describe effective non-equilibrium steady states \cite{bernard13}.

\section{NUMERICAL APPROACHES}
\label{sec:numerics}

Numerical computations in various classical and quantum dynamical transport problems in low dimensions, in particular in one-dimensional particle chains, have a rich and long history (see e.g. the reviews of \textcite{lepri} and \textcite{dhar}, and references therein). In this section we only mention classical and quantum molecular dynamics approaches which are suitable for efficient computer simulations. In the context of simulating coupled heat and matter transport from deterministic classical dynamical system's perspective one has to mention the references \textcite{mm2001} and \textcite{mm2003} which provided the first numerical measurements of the Onsager matrix for interacting chaotic classical gasses.

\subsection{Classical}
\label{sec:numerics_c}

The aforementioned and related studies simulate transport in an extended classical particle chain, or a particle container, by using a hybrid deterministic/stochastic method. The bulk dynamics is simulated deterministically using the standard techniques of molecular dynamics, i.e. solving Hamilton's equations for all particles' coordinates and momenta, while at the extreme (left and right) ends of the system, particles are exchanged with the left/right reservoirs (baths), or their energy is exchanged, in a stochastic fashion, so that after the event of stochastic interaction the particle's density and energy is distributed according to the grand-canonical distribution, determined by the temperatures and the chemical potentials of the baths. This can be viewed as a classical analog of the boundary driven open quantum many-body setup described in the previous section, and
can be implemented efficiently to estimate numerically the figure of merit $ZT$ \cite{casati08}.

There is also an alternative deterministic approach of the Nos\'e-Hoover thermostats \cite{nose}, in which also the dynamics of end (left or right) particles is purely deterministic, but it is dissipative and typically chaotic, so that these end particles are correctly thermalized (only for the average values, not for correlations).
Nos\'e-Hoover baths are nevertheless somewhat less used than the stochastic ones, as one can never be sure when deterministic dynamics of the baths may bring in some unwanted (spurious) correlation effects into the system's dynamics.

\subsection{Quantum}
\label{sec:numerics_q}

For quantum systems, boundary driven locally interacting Lindblad equation (\ref{eq:lind}) is particularly suitable since it allows for efficient numerical simulation of the steady state in terms of the time-dependent-density-matrix-renormalization-group method (tDMRG) \cite{daley,white,uli} in the Liouville space 
of linear operators acting on wave functions \cite{pz2009}. The reason for efficiency of this method in the long time limit as compared to the usual tDMRG algorithm lies in typically effective suppression of entanglement (correlations) in the operator space due to the coupling to Markovian baths. In the Liouvillean tDMRG approach, the state of the system, say of $N$ quantum spins $1/2$ or qubits (abstract two-level systems) described by Pauli matrices $\sigma^0\equiv \openone,\sigma^{1,2,3}\equiv \sigma^{x,y,z}$, encoded in a many-body density operator is at all times $t$ represented in the form of a {\em matrix product ansatz}, 
\begin{equation}
\rho(t) = \!\!\!\sum_{s_j\in\{0,1,2,3\}}\!\!\!\!\!\bra{L}A^{(s_1)}_1(t) \cdots A^{(s_N)}_N(t)\ket{R} \sigma^{s_1}\otimes\cdots\otimes\sigma^{s_N},
\label{eq:MPA}
\end{equation}
by means of a set of $4 n$ time-dependent matrices $A^{(s)}_j(t)$, and an appropriate pair of boundary vectors $\bra{L},\ket{R}$, of some finite dimension $D$.
The simplest strategy for calculating the density matrix of non-equilibrium steady state $\rho_\infty = \lim_{t\to\infty}\rho(t)$ is then simply to simulate time evolution of the master equation (\ref{eq:lind}), $\rho(t)=\exp(\hat{\cal L}t)\rho(0)$. Namely, $\exp(\hat{\cal L}t)$ can be decomposed into a product of local operators for systems with local interactions and local Lindblad dissipators $L_\mu$ which can be handled fully within the ansatz (\ref{eq:MPA}). The crucial approximation of the method lies in the fact that application of local two-site interaction operators $h_{[j,j+1]}$ on (\ref{eq:MPA}) results in amplification of dimension $D$ to $4D$, which in turn need to be truncated to $D$ by means of the singular value decomposition.
The overall error of such truncations is related to the growth of entanglement entropy (in the Hilbert-Schmidt space of density operators), which can be intuitively understood to be smaller for systems with dissipation as compared to fully coherent (Hamiltonian) evolution.
Having matrix product representation $\rho_\infty$ of the steady state one can then compute easily the local observables such as energy density, particle density or magnetization
profiles and currents from which the full Onsager matrix can be calculated.

One can also apply linear response approach and calculate Onsager coefficients from Green-Kubo expressions (\ref{eq:kubo}) based on tDMRG calculations of current-current 
time-correlation functions of pure Hamiltonian (coherent) dynamics. 
Here accessible time scales $t^*$, due to entanglement entropy growth, are much smaller than in Liouville space approach with dissipative boundaries, but $t^*$ does not need to be longer than the time up to which all current-current correlations essentially vanish. State-of-the-art algorithm for such calculations is described in \textcite{karrasch}.

In cases where the coupling among the particles is non-local, say we have residual long-range (e.g. Coulomb) interaction, or some other complications arise which prohibit efficient use of tDMRG, one can simulate quantum master equation using the method of quantum trajectories (see, for example, \textcite{mm2007}). The idea there is to represent the density operator as an expectation of 
$\ket{\Psi}\bra{\Psi}$ where the many-body wavefunction $\Psi$ is a solution of a stochastic Schr\"odinger equation $d\Psi(t) = -(i/\hbar) H\Psi dt + d\xi$, with $d\xi$ being an appropriate stochastic process simulating the action of the baths. The advantage of this technique is that non-Markovian effects can be treated easily and intuitively. 

For even more general problems, one can always solve the many-body quantum master equation numerically exactly, but the computational complexity then, of course, grows exponentially with the systems size. For modeling the proper canonical heat baths in quantum transport problems 
one often uses the Redfield equation 
\cite{redfield}.
Redfield master equation is derived physically via 
projection operator technique \cite{kubo}. 
It is a crucial property of the Redfield master equation 
that the steady state is 
guaranteed to have the Gibbs distribution.
Such approach has a wide applicabilities in many-body systems
and proved also useful to investigate thermal transport 
\cite{saito00,saito2003,segal05,wu08} and 
electron transport \cite{galperin09}.
A quantum Langevin equation approach was used to compute
thermal conductance through molecular wires \cite{hanggi2003}.

The time-dependent density functional theory was extended 
to include open quantum systems evolving under a master 
equation \cite{car,tempel} or in interaction with external baths
\cite{diventra07}. Applications of the density functional 
theory to thermoelectricity are discussed in \textcite{vignale,dagosta}.

\section{MODELS OF THERMODYNAMIC ENGINES}
\label{sec:machines}
Since the pioneering work by \textcite{carnot}, a huge number of physical phenomena have been 
recognized as heat engines, including thermoelectric phenomena. In original Carnot's idea, 
the maximum efficiency was studied for an ideal gas confined by a piston and 
is bounded by the Carnot efficiency.
The Carnot efficiency 
is achievable for the celebrated Carnot cycle, that consists of isothermal and 
adiabatic processes. Otto invented a more practical heat engine 
(Otto engine) \cite{callen} which
consists of adiabatic compression, heat addition at constant volume, 
adiabatic expansion, and rejection of heat at constant volume. 
In all heat engines
the maximum efficiency is obtained for a quasi-static process where asymptotically vanishing power output is generated. From the practical viewpoint, finite power with high efficiency is desirable.
This consideration opens the way to the concept of
finite-time thermodynamics, discussed in Sec.~\ref{sec:CA}.
Paradigms of finite-time thermodynamic engine are the Carnot or 
Otto cycles with a finite time period.
Recent technological developments enable us to consider and realize
finite-time thermodynamic devices with high controllability.
Main issues of finite-time thermodynamics 
are universal nature of efficiency at maximum power 
and new ideas for making such devices.
In this section, we discuss several paradigms of thermodynamic heat engines. 

\subsection{Stochastic thermodynamics}
\label{sec:stochastic}
Stochastic thermodynamics is a framework to study non-equilibrium thermodynamics in small 
systems like colloids or biomolecules driven out of equilibrium \cite{sekimoto, seifert}.
In most cases, this field treats classical systems.
The system is always attached to a thermal environment like water. 
The time scales 
of environment and system are sufficiently 
separated and the system's dynamics is well described by 
the Langevin equation. Suppose that one colloidal particle
is trapped by an external potential and the dynamics is overdamped;
the Langevin equation of motion is given by
\begin{eqnarray}
\dot{x} &=& \mu F ( x, \lambda (t) )  + \eta(t) \, ,
\end{eqnarray}
where $x(t)$ is the particle's coordinate,
$\eta (t)$ a Langevin thermal noise satisfying $\langle \eta (t ) \eta (t ' )
\rangle = 2D \delta (t-t') $ with
$D$ being the diffusion constant. The function
$F(x, \lambda (t) )$ is a time-dependent force field, which is given by
\begin{eqnarray}
F(x, \lambda (t)) &=& - \partial_x V(x, \lambda (t) ) + f\,, 
\end{eqnarray}
where $V(x,\lambda(t))$ is a potential and $f$ is 
a nonconservative force which can not be 
expressed as the gradient of a potential. 
The parameter $\lambda(t)$ is a time-dependent external
control parameter.
In equilibrium, the diffusion constant $D$ and the mobility 
$\mu$ are related by the 
Einstein relation $D=T\mu$.

The Langevin dynamics is endowed with a thermodynamic 
interpretation by applying the energy balance 
to any individual stochastic trajectory:
\begin{equation}
\delta U = \delta Q - \delta W,
\end{equation}
where $\delta Q$ is the heat absorbed from the environment, $\delta U$ 
the variation of internal energy, and $\delta W$ the work performed 
by the particle.
Note that $-\delta W$ is the work applied to the particle 
due to a time-dependent potential (i.e., $\lambda$ changes in time) 
or to a nonconservative force $f$. Since the dynamics is overdamped, 
the kinetic energy does not change in time.
Hence, the variation $\delta U$ of the internal energy 
is equivalent to the change $\delta V$ of the potential.  
The work applied to the particle reads
\begin{equation}
-\delta W= (\partial V/\partial \lambda) \dot{\lambda}dt + f dx,
\end{equation}
and therefore the 
absorbed heat is
\begin{equation}
\delta Q= \delta V + \delta W = -F dx \, .
\end{equation}
The work and 
heat defined above are the basis to consider thermodynamic 
efficiency in stochastic thermodynamics.

\begin{figure}
\begin{center}
\epsfxsize=75mm\epsffile{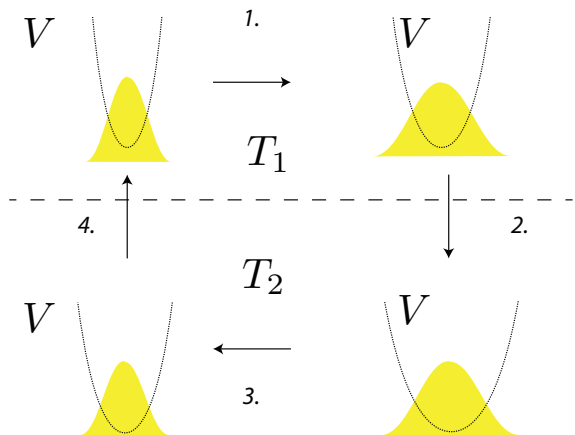}
\caption{(color online). Schematic picture of a stochastic thermodynamic engine.
In each plot the curve shows the potential $V$ versus $x$, 
the filled region is limited by the curve $p(x)$, representing
the (time-dependent) probability density to find the trapped
particle at $x$.}
\label{ss-engine}
\end{center}
\end{figure}


An intriguing and 
solvable heat engine in the framework of stochastic thermodynamics was introduced
by \textcite{ss08}. Suppose that one particle is trapped by a time-dependent 
harmonic potential $V(x, \lambda(t) ) = \lambda (t) x^2 /2$ 
without nonconservative force, i.e. $f =0$. 
We consider a cycle,
depicted in Fig.~\ref{ss-engine},
composed of the following four steps.
\begin{enumerate}
\item Isothermal transition at the hot 
temperature $T_1$ during $0 \le t < t_1$. 
The potential $V(x,\lambda(t))$ changes in time and work is extracted ($W>0$).
\item Adiabatic transition (instantaneously)
from the hot temperature $T_1$ to the cold temperature $T_2$.
\item Isothermal transition at the cold temperature $T_2$ during the time interval $t_1 \le t < t_1 + t_3$; $V(x,\lambda(t))$ changes in time and 
work is applied to the particle ($W<0$).
\item Adiabatic instantaneous transition from the cold temperature $T_2$ to the hot temperature $T_1$.
\end{enumerate}
This heat engine captures
important aspects of finite-time thermodynamics. 
Exact analysis of this model shows that the next leading order $\eta_C^2 /8$
in the CA formula is correct \cite{ss08}.
Later, this result was proved in a more general context \cite{esposito2009}.
The above described engine is quite realistic, since it
models the trapping of a colloidal particle in a 
time-dependent harmonic potential.
The stochastic heat engine was experimentally demonstrated \cite{bb2012}.

The CA efficiency was also studied in gas systems where 
the Carnot cycle is performed within a finite time cycle. 
The validity of CA efficiency was numerically discussed
by means of molecular dynamics simulations
\cite{izumida_okuda08,izumida_okuda091,izumida_okuda092}.

\subsection{Heat engine with blowtorch effect}
\label{sec:blowtorch}

\begin{figure}
\begin{center}
\epsfxsize=70mm\epsffile{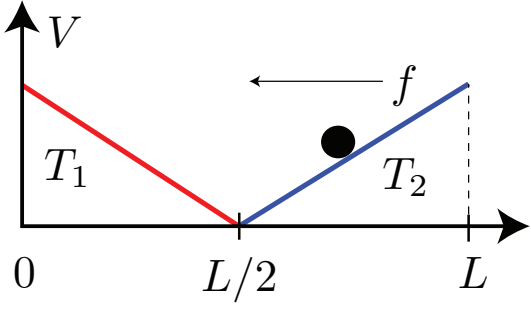}
\caption{(color online). Schematic picture of 
B\"{u}ttiker-Landauer heat engine.}
\label{blowtorch-engine}
\end{center}
\end{figure}

B\"{u}ttiker and Landauer's (BL) model is one 
example of heat engine \cite{buttiker87,landauer88,vankampen}. 
In BL model a particle is trapped
in a periodic potential $V(x)$
and subject to a nonuniform, spatially periodic temperature profile.
Although there is no time-dependent driving this system satisfies 
Curie's principle \cite{curie}, namely, rectification effect
occurs as it is not ruled out by symmetries.  
This situation is analyzed using the Langevin dynamics where 
the particle is alternately attached along the 
spatial coordinate, to thermal baths at different temperatures. The equation of motion 
is given by 
\begin{eqnarray}
m \ddot{x} &=& -\gamma(x) \dot{x} -V'(x)  -  f + \sqrt{2\gamma(x) T(x)} \, \xi (t)  , 
\end{eqnarray}
where $\gamma(x)$ is the coefficient of a viscous friction,
$f$ is the external force,
and $\xi $ is a white Gaussian noise satisfying 
$\langle \xi (t) \xi(t' )\rangle =\delta (t-t')$.
We assume that the potential and temperature depend on the position
and are periodic with period $L$, and take $(T(x), \gamma (x)) = 
(T_{1 (2)}, \gamma_{1 (2)})$
for $x \le (>) L/2$, with $T_1>T_2$. 
A schematic picture for the potential and temperature is presented in Fig.~\ref{blowtorch-engine}.

The energetics and transport properties of the BL motor have been studied by many authors. \textcite{landauer88} 
proposed the idea of the BL motor and showed 
the physical significance of nonuniform temperature in changing the relative 
stability of otherwise locally stable states, a phenomenon he termed as the blowtorch effect. 
Periodic temperature with a periodic potential induces a net transport of Brownian particles 
\cite{buttiker87,landauer88,parrondo02}. In the hot region 
the Brownian particle can move more easily than 
in the cold region. Hence, a finite net current is generated.
Efficiency $\eta$ is calculated by 
$\eta =\dot{W}/\dot{Q}_1$, 
where $\dot{W}$ is the work done by the particle per unit time:
$\dot{W}=f \langle \dot{x}\rangle$. The denominator $\dot{Q}_1$ is
the heat supply (per unit time) from the hot region, evaluated
as (see \textcite{sekimoto})  
$\dot{Q}_1 = \langle (-\gamma_1 \dot{x} + \sqrt{ 2\gamma_1 T_1} \xi (t)  ) \dot{x} \rangle$,
where the statistical average is taken only over the hot region, 
$0< x \le L/2$.

In the overdamped limit ($\ddot{x}\to 0$)
the net current can be derived using the Fokker-Planck equation. 
Then one can show 
that, under suitable conditions, the efficiency can reach 
the Carnot efficiency \cite{matsuo.sasa00,asfaw04,asfaw07}.
However, it was pointed out that reaching the Carnot efficiency is problematic 
due to the abrupt change of temperature at the boundaries
\cite{astumian99,hs00,ai05,ai06}. 
Carnot efficiency is unattainable due to the irreversible
heat flow via kinetic energy. 
Recent first principle calculations using molecular dynamics simulation 
support the unattainability of Carnot efficiency \cite{benjamin.kawai08}.

Brownian motors driven by temporal rather than spatial temperature
oscillations are discussed in \textcite{hanggiPLA}, where the 
potential has broken spatial symmetry (``ratchet'' potential).
The constructive role of Brownian motion for various physical 
and technological setups is reviewed
in \textcite{marchesoni}.

\subsection{Driven quantum dot}
\label{sec:drivenqd}
Among electric devices, the quantum dot (QD) has the potential to provide many kinds of 
thermodynamic engines \cite{linke2002,elb09,eklv10,linke2010}. 
Finite-time heat engines can be illustrated by means of quantum-dot 
systems, where one controls the gate voltage and in time to change the
on-site energy of the dot. 
We assume that a single-level quantum dot with time-dependent 
energy $\epsilon (t)$ is attached to one lead which has temperature $T (=1/\beta)$ and chemical potential 
$\mu (t)$.
Although the engine should work at the nanoscale, 
the dynamics is assumed to be classical, in that the
off-diagonal elements of the density matrix 
are neglected in the time-evolution.
Let $p(t)$ be the occupation probability for the state of the quantum dot. Then the time-evolution
is governed by the following master equation:
\begin{eqnarray}
\dot{p}(t) &=& - W_1 (t) p(t) + W_2 (t) \left[ 1 - p(t) \right] \, ,
\end{eqnarray}
where the transition rates read
\begin{eqnarray}
\begin{array}{ll}
W_1 (t) = {C \Bigl[  e^{-\beta \left[ \epsilon (t) - \mu (t) \right]} + 1}\Bigr]^{-1} \, ,\\
W_2 (t) = {C \Bigl[ e^{\beta \left[ \epsilon (t) - \mu (t) \right]} + 1} \Bigr]^{-1} \, ,
\end{array}
\end{eqnarray}
with a constant $C$. The internal energy $U(t)$ of the quantum-dot system 
at time $t$ is defined as
$U(t)=\varepsilon(t) p(t) $; 
the output work $\delta W(t)$ and the absorbed heat 
$\delta Q(t)$ are given by 
\begin{eqnarray}
\delta W (t)&=& -\left[ d \epsilon (t) - d \mu (t)  \right] p(t)  \, , \\
\delta Q (t)&=& \left[ \epsilon (t) - \mu (t)  \right] d p(t)  \, .
\end{eqnarray}

\begin{figure}
\begin{center}
\epsfxsize=75mm\epsffile{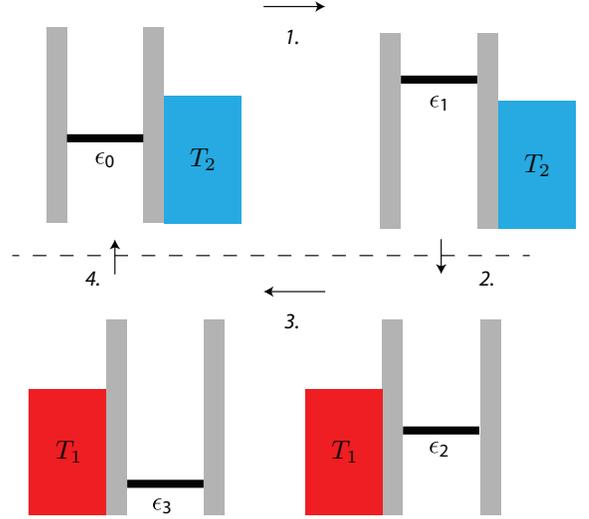}
\caption{(color online). Schematic picture of a heat engine 
that consists of a driven quantum dot.}
\label{dot-engine}
\end{center}
\end{figure}

\textcite{eklv10} proposed a solvable engine using 
the following cycle (see Fig.~\ref{dot-engine}):
\begin{enumerate}
\item Isothermal process: The 
quantum dot is in contact with a cold lead at temperature $T_2$ and chemical 
potential $\mu_2$. The energy is raised during a finite time $\tau_a$ 
as $\epsilon (t): \epsilon_0 \to \epsilon_1 ~~(\epsilon_1 > \epsilon_0)$.
\item Adiabatic process: The quantum dot is disconnected from the lead, and the energy is abruptly lowered
as $\epsilon (t): \epsilon_1 \to \epsilon_2 ~~(\epsilon_1 > \epsilon_2)$.
\item Isothermal process: The quantum dot is connected to a hot lead with temperature $T_1$ and chemical 
potential $\mu_1$. The energy is lowered during a finite time $\tau_b$,
$\epsilon (t): \epsilon_2 \to \epsilon_3 ~~(\epsilon_2 > \epsilon_3)$.
\item Adiabatic process: The dot is disconnected, and the energy abruptly returns to the original value:
$\epsilon (t): \epsilon_3 \to \epsilon_0$.
\end{enumerate}
The period of one cycle is $\tau=\tau_a + \tau_b$, the output power 
is given by 
$P=W/ \tau = Q/ \tau =\int_{0}^{\tau} dt \, \dot{p}(t) \left[ \varepsilon(t) - \mu (t)\right] $.
Finding the set of parameters that maximize the power 
may be done with a variational equation.
In particular, the CA efficiency is recovered in the limit of weak dissipation
\cite{eklv10}. 

\subsection{Quantum heat engines}
Quantum mechanics and thermodynamics have a deep connection,
whose investigation started from 
thermodynamic studies by \textcite{planck01} and \textcite{einstein17}.
Due to recent progress of micro-fabrication technology, 
quantum effects in small heat engines 
have become an important subject. 

The expectation value of the measured energy of a whole quantum system is $U=\sum_i p_i E_i$, 
where $E_i$ are the energy levels and $p_i$ are the corresponding occupation probabilities.
This implies that  
\begin{eqnarray}
dU &=&\sum_i \left(E_i dp_i + p_i dE_i\right) \, ,
\end{eqnarray}
from which the absorbed heat
is recognized to be $\delta Q=\sum_i E_i dp_i$ and the output 
work $\delta  W =-\sum_i p_i d E_i$. 
Then, the first law $dU=\delta Q-\delta W$ is satisfied.

The quantum-mechanical analogue of Carnot cyclic engine is introduced by identifying
isothermal processes with quantum processes at constant expectation value
of the energy \cite{bender00}. 
For a single quantum mechanical particle confined to a potential well, 
the Carnot efficiency in this framework takes the form 
$\eta_C = 1- {E_2 / E_1}$, where 
$E_1$ and $E_2$ are the energy expectation values
during the isothermal transformations at temperatures
$T_1$ and $T_2$, respectively ($T_1>T_2$). 
Entropy and temperature were discussed within this framework \cite{bender02}. 
Extensions of this approach include ideal Fermi gas
with an arbitrary number of particles \cite{wang12},
Otto cycle \cite{wu10},and the investigation of the efficiency at 
maximum power for the engine introduced by \cite{abe11}.

Quantum version of the Carnot heat engine and Otto heat engine for
finite temperature was defined without ambiguities in 
\cite{quan07}. Effect of multilevel systems was invesitigated \cite{quan05}. 
A class of quantum heat engines consisting of two subsystems interacting with a work source was studied 
to maximize the extracted work under various constraints \cite{mahler08}.
Concepts from quantum information theory 
also provide new insights into the working of 
quantum heat engines \cite{quan06,zhou10,nori09,kieu06,kieu04}. 

Open systems attached to a thermal environment
can be analyzed by means of the 
quantum master equation in the form \cite{lindblad,redfield,kubo}
\begin{eqnarray}
\dot{\rho} &=& \frac{i}{\hbar} \Bigl[ \rho ,  H(\lambda (t))  \Bigr] -\sum_{\alpha} \hat{\Gamma}_{\alpha} (t) \, \rho(t) 
\, ,
\end{eqnarray}
where $\rho$ is the density matrix of the system and $\hat{\Gamma}_{\alpha} (t)$ denotes
a superoperator representing dissipative effects arising from a reservoir labeled by index $\alpha$.
The parameter $\lambda$ is a control parameter for the system. 
Then the output work $\delta W$ and the 
heat $\delta Q_{\alpha}$ absorbed from the $\alpha$ reservoir are given by
\begin{eqnarray}
\delta W &=& - {\rm Tr} \left[ \, \rho (t) \partial_{\lambda} H (\lambda) \, \right] \dot{\lambda} dt \, ,
\\
\delta Q_{\alpha} 
&=& -{\rm Tr} \left[ H (\lambda) \hat{\Gamma}_{\alpha} (t)  \rho (t) \right] dt \, .
\end{eqnarray}
The master equation approach reproduces the Carnot efficiency for the Carnot cycle \cite{alicki79}.
The performance of a quantum heat engine 
or heat pump with the working fluid
composed of noninteracting two-level systems 
was investigated by using a master equation \cite{feldmann00}.
\textcite{kosloff84} showed that
two coupled oscillators interacting with hot and cold quantum 
reservoir exhibit Curzon-Ahlborn efficiency in 
the limit of weak coupling.
The quantum master equation was applied to analyze the
performance of heat engines working with  
spins \cite{geva92,he02,chen02}, harmonic oscillators 
\cite{lin03-1,amaud02, lin03-2, lin03-3},
and multi-level systems \cite{geva94}.
Characteristics of the steady state 
achieved by the iteration of cyclic process and monotonic approach to the limit cycle was discussed 
making use of the quantum conditional entropy \cite{feldmann04}.
The unavoidable irreversible loss of power in a heat engine
was considered for harmonic systems in the framework of quantum master 
equation approach \cite{rezek06}.

Fifty years after the pioneering works
on quantum mechanics and thermodynamics by 
\textcite{planck01} and \textcite{einstein17}, models of lasers and masers were realized 
as quantum heat engines \cite{scovil59} and the relation between the quantum efficiency of the maser and
the Carnot cycle was investigated. Detailed balance imposed by thermodynamics 
limits the efficiency 
of quantum heat engines, including solar cells \cite{shockley61}.
Scully {\em et al.} analyzed the performance of a quantum heat
engine operating 
by means of 
the radiation pressure from a single mode radiation field which drives
a piston engine or a photon-Carnot engine. They pointed out that the phase associated with
the atomic coherence provides a new control parameter, which can be varied to increase the
temperature of the radiation field and to extract work from a single heat bath, while the real
physics behind the second law of thermodynamics is not violated \cite{scully02,scully03}.
This photon-Carnot engine was discussed in a
quantum information context \cite{lutz09} and 
decoherence mechanisms were studied \cite{quan06-2}.

Breakdown of detailed balance due to quantum coherence has been discussed for 
a quantum-dot photocell \cite{scully10} 
and a photon-Carnot quantum heat engine 
\cite{scully03}.
Noise-induced coherence was proposed as a potential ingredient to 
enhance quantum power \cite{scully11}.
Quantum Otto engine in two-atom cavity quantum electrodynamics \cite{wang09} 
was studied to understand the role of thermal entanglement,
and realization of Otto engine in single-modes fields was discussed 
\cite{wang12-2}.

Photosynthetic reaction center 
is a quantum engine where energy is supplied from the light
and hence it has similarities to heat engine \cite{dorfman13}.  
This also provides 
an important intersection between physics and biology.
The rich history and ongoing studies suggest that the deep underlying connection between thermodynamics 
and quantum mechanics may be useful 
for improving the design and boosting the efficiencies
of light-harvesting devices. 

The concept of ideal quantum heat engine was introduced in 
cold bosonic atoms confined to a double well potential 
where thermalization occurs,
and operation of a heat engine with a finite quantum heat bath was 
theoretically demonstrated \cite{fialko12}.
A thermoelectric heat engine with ultracold fermionic atoms
was demonstrated, both theoretically and 
experimentally \cite{esslinger}.

The fundamental limits to the dimensions of 
(steady-state) self-contained quantum machines, 
i.e. machines working without the supply
of external work but only due to interactions with thermal
baths at various temperatures, were investigated by 
\textcite{popescu2010,popescu2011,popescu2012}. 
In particular, it was shown that also a small self-contained 
refrigerator consisting of three qubits, each one in
contact with a thermal reservoir, can achieve the Carnot 
efficiency \cite{popescu2011}. In \textcite{fazio2013}
an electronic quantum refrigerator based on four quantum 
dots in contact with as many thermal reservoirs
was theoretically investigated.

\section{PHENOMENOLOGICAL LAWS AND THERMOELECTRIC TRANSPORT}
\label{sec:phenomenology}

\subsection{Wiedemann-Franz law}
\label{sec:Wiede}

In a wide range of macroscopic electronic conduction, the electron conductivity
and the electronic contribution to
the thermal conductivity are not independent.
They are connected to each other by an empirical relation called
the Wiedemann-Franz (WF) law \cite{wf}. The WF law states that the
ratio of the 
thermal conductivity $\kappa$ to the electric conductivity
$\sigma$ of a metal is proportional to the temperature:
\begin{equation}
\frac{\kappa}{\sigma} = \mathfrak{L} T \, ,  \label{wflaw}
\end{equation}
where the constant value $\mathfrak{L}$ is known as the
Lorenz number. In non-interacting electronic systems at low temperatures,
the Lorenz number is given by
\begin{equation}
\mathfrak{L} = \frac{\pi^2}{3} \left( \frac{k_B}{e} \right)^2 . \label{lorenz}
\end{equation}
This law is derived by using either kinetic theory or Landauer formula.
In the kinetic theory approach \cite{kittel}, one uses the expression of
electronic conductivity $\sigma \sim n e^2 \tau / m$,
where $n$ is the electronic density and $\tau$ is the mean-free time.
Thermal conductivity $\kappa$ is given by
$\kappa \sim C v_F^2 \tau/3$, with $C$ the specific heat
and $v_F$ the Fermi velocity. The specific heat for a free electronic
system at low temperatures is given by
$C \sim \pi^2 n k_B^2 T / m v_F^2$, which immediately yields the constant
value (\ref{lorenz}) for the Lorenz number.
Another approach is based on the Landauer's expressions of 
electric and thermal conductance, 
Eq.~(\ref{eq:landauer.ex}). In these expressions, we assume 
that the transmission is weakly energy-dependent, hence
\begin{equation}
\tau (E) \approx \tau (\mu) \, .
\end{equation}
Under this assumption, one can derive the WF law for the ratio
of thermal and electric conductances:
$\Xi/G=\mathfrak{L}T$.
Both derivations of the WF law are substantiated by
the Sommerfeld expansion \cite{ashcroftmermin} of integrals
(\ref{eq:In}) or (\ref{eq:Kn})
to the lowest order in $k_B T/E_F$, with $E_F$ being the Fermi energy.
Such expansion is valid for a smooth function $\tau(E)$ or
$\Sigma(E)$. Note that to derive the WF law the off-diagonal
Onsager coefficient $L_{12}$ has to be neglected, i.e.,
one needs $L_{11}L_{22}\gg L_{12}^2$ in
order to approximate the thermal conductance 
$\Xi$ with $L_{22}/T^2$ \cite{beenakker1992}.

\subsection{Mott's formula}
\label{sec:Mott}

The Mott's formula states that the Seebeck coefficient
$S$ in a metal is approximately given by the
formula \cite{mott1,mott2,mott3}
\begin{equation}
S \sim
\frac{\pi^2 k_B^2 T}{3 e} \partial_{E} \ln G (E) \Bigr|_{E=\mu} \, , \label{mott}
\end{equation}
where $G (E)$ is the electric conductance at chemical potential $E$ in the leads.
This relation is obtained
under the assumption that the system is non-interacting and
that conduction mainly occurs around the Fermi energy.
Thus, the transmission probability is approximated as follows:
\begin{equation}
\tau (E) \approx \tau (\mu) + \partial_E \tau (\mu) \, (E-\mu )  \, .
\end{equation}
The Mott's relation (\ref{mott}) is derived after inserting this expansion
into (\ref{eq:landauer.ex}).
An analogous derivation is obtained by
considering the conductivity
$\sigma(E)$ rather than conductance $G(E)$ and
(\ref{eq:boltzmann.ex}) rather than (\ref{eq:landauer.ex}).

\subsection{Thermoelectricity}
\label{sec:thermoelectricity}

The Mott's formula tells us that
a sharp energy-dependence of electric conductivity is crucial for
increasing the Seebeck effect. This is consistent
with the importance of energy filtering structure of
transmission discussed in Sec.~\ref{sec:energyfiltering}.
Both the WF law and the Mott's relation follow
from the  single particle Fermi-liquid (FL) theory, so that the
Sommerfeld expansion can be applied. If the FL theory
holds in non-interacting systems, the WF law is valid in the
presence of arbitrary disorder \cite{chester,jm80}.
When the WF law is valid, it is not possible to obtain large
thermoelectric efficiency.
Indeed, the WF law is obtained under the condition
$L_{11}L_{22}\gg L_{12}^2$, hence the figure of
merit $ZT=L_{12}^2/{\det{\bm L}}\approx L_{12}^2/L_{11}L_{22}\ll 1$.
This implies that to get large $ZT$ one should search for physical situations where the WF law is violated.
The WF law is largely violated in low-dimensional interacting systems
that exhibit non-FL behavior
\cite{kf96,lo02,dora06,grsr09,wbxmgh11} and in small systems where
transmission can show a significant energy dependence
\cite{fazio2001,stone,bss10,bjs10}. 
Properties of reservoirs for such small systems may affect the
validity regimes of the WF law \cite{bbb12}.
Departures of the WF law in the regime of nonlinear transport 
were investigated in \textcite{sanchezprb13}. 
The possibility of engineering the Lorenz number in 
superlattices was discussed in \textcite{shakouri2007b}.

Hicks and Dresselhaus theoretically studied quantum-well
structures in low dimensions and showed that layering
has the potential to increase the thermoelectric figure
of merit \cite{hd93,hd93b}. Indeed the layering may reduce the 
phonon thermal conductivity as phonons can be scattered  
by the interfaces between layers. Moreover, sharp features
in the electronic density of states, favorable for
thermoelectric conversion
(see the discussion of Sec.~\ref{sec:energyfiltering}), 
are in principle possible
due to quantum confinement.
This proposal was experimentally demonstrated \cite{hhsd96}.
Moreover, it motivated interest in the thermoelectric properties
of low-dimensional nanostructures.
For experimental evidence of energy filtering, see 
\textcite{shakouri2006}, where barriers in a
superlattice were used to limit the transport to those electrons
with sufficiently high energy. As a result, a dramatic increase of
the Seebeck coefficient was shown together with a relatively modest 
decrease of the electrical conductivity.
Significant energy-dependence of transmission was found in graphene
\cite{zuev09}. 
Thermopower measurements have been used to study
semiconductor nanostructures such as quantum point contacts \cite{mhbef90}
and quantum dots in the Coulomb blockade regime \cite{molenkamp94,staring93}.
The thermoelectric properties of a quantum dot can be used to obtain
thermal rectification, as reported in the experiments by 
\textcite{molenkamp2008}.
First principle calculation for silicon nanowires shows that 
in nanowires
small diameter is desired to increase $ZT$ and that isotopic doping 
can increase the value of $ZT$ significantly \cite{shi09}.
Sharp electronic resonances can be found also in molecules 
weakly coupled to electrodes and for this reasons molecular 
junctions might be efficient for thermoelectric conversion
\cite{majumdar2007,reddy2012,pauly2008,baranger2009,lambert2009,lejinse2010,bss10,dubi,venkataraman2013}. 
Photo-Seebeck effect, i.e. the contribution of photo-induced 
carriers to the thermopower, was experimentally demonstrated 
for a wide-gap oxide semiconductor \cite{okazaki12}.

Small systems such as quantum dots have an advantage 
for making narrow energy window in 
transmission and breaking the WF law. 
On the other hand, small systems exhibit large 
fluctuations in physical quantities.
In open chaotic quantum dots, thermopower shows significant fluctuations.
The distribution of parametric conductance derivatives was analyzed
using random matrix theory \cite{brouwer97} and fluctuations
of thermopower were discussed in \textcite{lsb98}.
Thermopower fluctuations exhibit a non-Gaussian distribution,
experimentally demonstrated in \textcite{gbm99}.
Effects of the spectrum edges of a chaotic scatterer on 
thermopower distributions and its universal 
aspects were studied in \textcite{pichard13}. 
Fluctuations and nonlinear response of 
thermoelectricity were studied in terms of 
time-reversal symmetry \cite{iyoda10} in analogy 
to the expansion using the fluctuation 
theorem for electric transport \cite{saito08,andrieux09}.
Nernst effect was theoretically studied to get strict bounds on efficiency
\cite{stark}.

\section{CONCLUSIONS}
\label{sec:conclusions}

In this Colloquium we have presented a simple and self-contained account of few main theoretical approaches to discuss the problem of thermoelectric efficiency and the efficiency of steady-state heat to work conversion in general. Even though the problem has a long history, we believe that many recent theoretical advances described here, being in turn stimulated by new generations of experiments with nano-scale systems, should be somewhat taken from a new, more abstract perspective. Namely, we believe that the powerful machinery of non-equilibrium statistical mechanics and dynamical system's theory has not yet been fully explored in connection to coupled heat and electric, magnetic or particle transport and in particular in analyzing the figure of merit of thermoelectric, thermomagnetic or thermochemical efficiency. This being particularly so in view of some very recent new fundamental results on the behavior of thermoelectric efficiency in the presence of time-reversal symmetry breaking of the underlying equations of motion, such as by means of the magnetic field \cite{BSC2011,BSS2013,BS2013}.

The central question which identifies this paper is:  What limits, if any, the microscopic dynamical laws -- for a particular model, or for a particular non-equilibrium steady-state setup -- 
impose on the thermodynamic heat-to-work efficiency?
While the theory for non-interacting systems discussed here seems to be well understood, the understanding of general mechanisms connected to strong interactions are only beginning to emerge. This justifies the importance of efficient numerical simulations of interacting systems at this point.
While this Colloquium is focused on linear transport, thermoelectric devices
often operate in the nonlinear regime of transport. In that regime,
reciprocity relations break down, as experimentally observed  
in a four-terminal mesoscopic device \cite{matthews13},
and the figure of merit $ZT$ fails 
to describe thermoelectric performance \cite{shakouri2007,jacquod13,whitney13b}.
The observed breakdown of the Onsager-Casimir relations
increasing thermal bias could in principle allow for improved
thermoelectric efficiencies.
In the nonlinear regime, 
rectification effects occur and their impact on thermoelectricity
is still not well understood.   
In this regard, the recently developed \cite{sanchez13,jacquod13,whitney13}
scattering theory of nonlinear thermoelectricity could pave the way to 
deeper investigations of quantum coherent conductors 
working beyond the linear response regime.

We expect to see in near future a burst of applications of fundamental ideas on abstract dynamical mechanisms for analyzing, or engineering particular practically relevant models, or for designing experiments and technological applications.

\section*{Acknowledgments}

G.B. and G.C. acknowledge the support by MIUR-PRIN.
T.P. acknowledges support by the grants P1-0044 and J1-5439 of Slovenian 
Research Agency (ARRS).
K.S. was supported by MEXT (23740289) .

\bibliographystyle{apsrmp}


\end{document}